\documentclass[11pt]{article}
\usepackage[margin=0.625in]{geometry}
\usepackage[numbers]{natbib}   %

\global\bibsep=0pt
\usepackage{mathtools}
\usepackage{amsfonts}
\usepackage{amssymb}
\usepackage{amsthm}
\usepackage{soul}
\usepackage[dvipsnames]{xcolor}
\usepackage{algorithm,algpseudocode}
\usepackage{stmaryrd}
\usepackage{caption}
\usepackage{hyperref}
\hypersetup{
    colorlinks,
    linkcolor={red!50!black},
    citecolor={blue!50!black},
    urlcolor={blue!70!black}
}

\usepackage{newfloat}
\DeclareFloatingEnvironment[
  fileext = lom,
  listname = {List of movies} ,
  name = Mov.
]{movie}

\newcommand{\coeff}[1]{\hat {#1}}
\newcommand{\pair}[1]{\llbracket #1 \rrbracket}
\newcommand{\norm}[1]{\| #1 \|}

\algnewcommand{\Inputs}[1]{%
  \State \textbf{Inputs:}
  \Statex \hspace*{\algorithmicindent}\parbox[t]{.8\linewidth}{\raggedright #1}
}
\algnewcommand{\Initialize}[1]{%
  \State \textbf{Initialize:}
  \Statex \hspace*{\algorithmicindent}\parbox[t]{.8\linewidth}{\raggedright #1}
}

\def\Balpha{\boldsymbol{\alpha}}

\def\Bsigma{\boldsymbol{\sigma}}

\def\Bphi{\boldsymbol{\phi}}

\def\Bpsi{\boldsymbol{\psi}}
\def\Bomega{\boldsymbol{\omega}}
\def\Bvarepsilon{\boldsymbol{\varepsilon}}

\def\bE{\boldsymbol{E}}
\def\bF{\boldsymbol{F}}
\def\bG{\boldsymbol{G}}
\def\bH{\boldsymbol{H}}
\def\bI{\boldsymbol{I}}

\def\bQ{\boldsymbol{Q}}
\def\bR{\boldsymbol{R}}

\def\bU{\boldsymbol{U}}

\def\bW{\boldsymbol{W}}

\def\be{\boldsymbol{e}}
\def\bff{\boldsymbol{f}}
\def\bg{\boldsymbol{g}}

\def\bl{\boldsymbol{l}}

\def\bn{\boldsymbol{n}}

\def\bp{\boldsymbol{p}}
\def\bq{\boldsymbol{q}}
\def\br{\boldsymbol{r}}

\def\bu{\boldsymbol{u}}
\def\bv{\boldsymbol{v}}
\def\bw{\boldsymbol{w}}

\def\Base{\boldsymbol{\mathrm{1}}}
\def\Img{\boldsymbol{\mathrm{i}}}
\def\Complex{\mathbb{C}}
\def\Real{\mathbb{R}}

\def\Lie{\mathcal{L}}

\def\cvr{\nabla}
\def\blp{\Delta}
\def\grad{\mathrm{grad}}
\def\div{\mathrm{div}}
\def\delbar{\bar \partial^{\cvr}}

\DeclareMathOperator{\rot}{rot}
\DeclareMathOperator{\symm}{symm}

\newcommand{\Pe}{\mbox{\em Pe}}

\newcommand{\thinsim}{{\raise.17ex\hbox{\(\scriptstyle\mathtt{\sim}\)}}}

\makeatletter
\newsavebox{\@brx}
\newcommand{\llangle}[1][]{\savebox{\@brx}{\(\m@th{#1\langle}\)}%
  \mathopen{\copy\@brx\kern-0.5\wd\@brx\usebox{\@brx}}}
\newcommand{\rrangle}[1][]{\savebox{\@brx}{\(\m@th{#1\rangle}\)}%
  \mathclose{\copy\@brx\kern-0.5\wd\@brx\usebox{\@brx}}}
\makeatother
\def\cf{cf.~}
\def\eg{e.g.,~}

\def\ie{i.e.,~}

\begin{document}

\begin{centering}
    \textbf{\Large Active nematic fluids on Riemannian 2-manifolds}\\[3mm]
    \textbf{Cuncheng Zhu\footnote{Department of Mechanical and Aerospace Engineering, University of California San Diego, USA, cuzhu@ucsd.edu.}, David Saintillan\footnote{Department of Mechanical and Aerospace Engineering, University of California San Diego, USA, dstn@ucsd.edu.}, Albert Chern\footnote{Department of Computer Science and Engineering, University of California San Diego, USA, alchern@ucsd.edu.}}\\[1mm]
\end{centering}

\begin{abstract}
    Recent advances in cell biology and experimental techniques using reconstituted cell extracts have generated significant interest in understanding how geometry and topology influence active fluid dynamics.
    In this work, we present a comprehensive continuous theory and computational method to explore the dynamics of active nematic fluids on arbitrary surfaces without topological constraints.
    The fluid velocity and nematic order parameter are represented as the sections of the complex line bundle of a 2-manifold.
    We introduce the Levi-Civita connection and surface curvature form within the framework of complex line bundles.
    By adopting this geometric approach, we introduce a gauge-invariant discretization method that preserves the continuous local-to-global theorems in differential geometry.
    We establish a nematic Laplacian on complex functions that can accommodate fractional topological charges through the covariant derivative on the complex nematic representation.
    We formulate advection of the nematic field based on a unifying definition of the Lie derivative, resulting in a stable geometric semi-Lagrangian discretization scheme for transport by the flow.
    In general, the proposed surface-based method offers an efficient and stable means to investigate the influence of local curvature and global topology on the 2D hydrodynamics of active nematic systems.
    Moreover, the complex line representation of the nematic field and the unifying Lie advection present a systematic approach for generalizing our method to active $k$-atic systems. 
\end{abstract}

\section{Introduction}\label{sec: introduction}

Active nematics, a unique class of systems defined by the interaction between intrinsic activity and orientational order, have emerged as a captivating  paradigm to explore complex dynamical phenomena across scales.
In biophysical contexts, deciphering the dynamics of these active matter systems offers profound insights into a myriad of processes, encompassing collective dynamics in motile bacterial suspensions, structural organization in cytoskeletal assemblies, the dynamics of self-propelled colloidal particles, and the processes governing tissue morphogenesis \cite{alert2022active, doostmohammadi_active_2018}.
Recent experimental discoveries and the potential for significant advances in fields like material science and biophysics have sparked a growing interest in understanding the intricate connections between the dynamic behavior of active nematic systems and the geometric and topological constraints imposed by spatial confinement \cite{firouznia2024self,apaza2018active, mietke_self-organized_2019, hoffmann_theory_2022, bell2022active, al-izzi_active_2021, al-izzi_morphodynamics_2022, zhang2016dynamic, keber_topology_2014, napoli_extrinsic_2012, napoli2016hydrodynamic}.
However, the theoretical understanding and technological advancements in studying these systems are currently limited. 
Mathematically, there exists a fundamental connection between tensor fields describing the nematic system with the underlying geometry and topological characteristics of the space it lives in. 
This relationship can formally described by the mathematical concept known as a fiber bundle structure.
In this paper, we introduce active nematic systems situated on generalized Riemannian surfaces using a geometric language and the complex line bundle formalism.  
Our focus is on understanding how nematic order parameters relate to the intrinsic curvature and global topological properties of the underlying space.
We also present a straightforward and stable computational methodology that aligns with this theoretical framework. 
We expect the method to be a useful tool for practitioners studying active nematic systems where the local curvature and global topology of the system is important, such as in morphogenesis, and for exploring potential engineering applications in soft robotics and morphing materials. 

To characterize nematic fields, the prevailing approach in the existing literature relies on the Landau-de Gennes $\boldsymbol{Q}$-tensor theory, which describes the configuration of the nematic field on a surface via a second-order symmetric tensor. 
However, in 2D there exists an alternative representation through a homogeneous quadratic complex scalar function.
This approach offers distinct advantages because of the inherent constraints on its degrees of freedom.
Unlike the complex representation, the evolution of the $\boldsymbol{Q}$-tensor needs to be confined within the realm of traceless and symmetric matrices, which necessitates additional complications such as the use of Lagrange multipliers during evolution \cite{sonnet_dissipative_2012, salbreux2009hydrodynamics}. 
Furthermore, the complex representation facilitates a natural extension to $k$-atic fields (\ie director fields with $k$-rotational symmetry) by simply raising the exponent of the complex function to $k$.
Recently, there has been significant interest in modeling tissue dynamics through $k$-atic liquid crystal theory, highlighting the practical relevance of this approach \cite{giomi2022hydrodynamic}.

The mathematical formulation of nematohydrodynamics, or more broadly of complex fluid flows, has been extensively explored in the literature \cite{doiedwards1988,Bird1988,berisedwards1994,saintillan2013active,morozov2015introduction}. 
Classic approaches primarily rely on an explicit approach, employing coordinate-based expressions and standard vector calculus. 
They often focus on the mechanics of specific physical objects in flow and root their arguments in the principle of material frame invariance \cite{morozov2015introduction}.
These approaches become particularly challenging when attempting to translate ordinary vector calculus identities to curved surfaces.
However, in the realm of differential geometry, there exists an abstract yet powerful geometric language, exemplified by the Lie derivative, which functions independently of coordinate systems. 
This language offers a natural way to comprehend various physical phenomena and conservation laws in fluid dynamics \cite{gilbert_geometric_2023}. 
Concepts such as convected time derivatives in complex fluids can be effectively generalized through the use of the Lie derivative, unifying the derivation process.
Despite its generality and utility, the adoption of this geometric language remains relatively obscure in applied fluid mechanics and is not widely recognized in the field of active nematics. 
In light of these challenges, our objective is to bridge this knowledge gap and recast a minimal active nematic system using the Lie derivative.
This effort not only provides a unified perspective on the various components of the active nematohydrodynamic system but also, as we will demonstrate in the methods section, leads to a stable and elegant numerical approach \cite{wensch_semi-lagrangian_2012, nabizadeh_covector_2022, mullen_discrete_2011}.

Similar to the Lie derivative, we also seek to establish a coordinate-free framework for analyzing active nematic fields on curved surfaces by using the mathematical structure called \emph{fiber bundles}. 
This structure encapsulates the concept of a family of tensor spaces parameterized by a base manifold, providing a coordinate-free geometric language for describing the tensor fields on the manifold as \emph{sections} of bundles. 
When each \emph{fiber} in the bundle is a one-dimensional complex vector space equipped with a Riemannian metric, which houses the $k$-atic director and velocity vector, this bundle is considered a Hermitian complex line bundle. 
The theory of the complex line bundle has been the subject of extensive research in both pure mathematics and theoretical physics, particularly in modeling condensed matter phenomena \cite{Pismen1999VorticesIN}. 
The Levi-Civita connection, or covariant derivative, connects each fiber within the tangent bundle. 
The curvature of the base manifold manifests from the holonomy of this connection. 
The Levi-Civita connection on the tangent bundle induces an associated connection onto the nematic bundle, which gives rise to a nematic Bochner Laplacian that accommodates fractional topological defects under the relaxation of nematic elasticity.
The nuanced difference between the nematic and vector Bochner Laplacians has often been overlooked or inadequately explored in the existing literature.
In addition to the theoretical advancement it provides, this complex line representation also facilitates the construction of an intrinsic complex-valued discrete Laplacian matrix based on the local chart, instead of relying on a global parametrization or an $\Real^3$ representation in the embedded coordinate. 
Overall, the formalism grounded in the complex line bundle theory unveils a richer perspective on the representation of nematic---and more broadly $k$-atic---fields, across arbitrary 2-dimensional manifolds.

We also show how one can formulate the Stokes equations through the Dolbeault operators in the complex line bundle theory.
In the smooth setting, the flow of a viscous medium follows the Onsager variational principle. 
According to this principle, the velocity of the viscous film minimizes a dissipation measure known as the Rayleighian, which is a function of the strain rate tensor.
On a Riemannian manifold, the measurement of strain rate is determined by the symmetric part of the covariant derivative, commonly referred to as the Killing operator.
With the complex structure, the covariant derivative can be orthogonally decomposed into the holomorphic \emph{Del} and the anti-holomorphic \emph{Del bar} operators, jointly known as the Dolbeault operators.
We elucidate that, under the incompressibility condition, the Laplacian-like operator that determines the viscous stress on a Riemannian manifold is induced by the Del bar operator.
In addition to the Bochner Laplacian, this operator encompasses contributions from the curvature of the manifold.
This fact is acknowledged in the literature, despite the results having been derived from the perspective of Riemannian geometry \cite{gilbert_geometric_2023, chan_formulation_2017, samavaki_navierstokes_2020}.
While both derivations are equivalent in this specific case, we believe that the more general formulation based on complex line bundles can reveal new perspectives and potential for further theoretical or computational advancements.

In the computational realm, several efforts have previously been undertaken to solve active fluid dynamics on curved surfaces.
These endeavors are primarily founded on continuum formulations \cite{rank_active_2021, brandner_finite_2022, reuther_numerical_2020}  or on the Lattice Boltzmann method \cite{zhang2016lattice, zhang2016dynamic}. 
In particular, we highlight the limitations of using the streamfunction formulation for incompressible fluid mechanics on manifolds with non-trivial topology without specialized modifications \cite{nitschke_vorticitystream_2021, yin_cohomology, reusken2020stream, pearce_geometrical_2019, torres-sanchez_modelling_2019, brandner_finite_2022}. 
However, one common challenge with velocity-pressure-based methods is the representation of tangent vectors. 
In our method, we represent both the velocity and director using an intrinsic complex-valued function alongside the Levi-Civita connection. 
This approach eliminates the need for the commonly used numerical technique of constraining the $\mathbb{R}^3$-valued velocity to lie tangent to the manifold \cite{brandner_finite_2022, gross2018trace}.
Moreover, we draw parallels between the numerical scheme for active nematics and classic problems in computational geometry.
For example, insightful analogies can be found from the resemblance between the Stokes equations and the computation of Killing vector fields, which aid in discerning symmetries within geometries \cite{ben-chen_discrete_2010, de_goes_discrete_2014, de2020discrete, knoppel_globally_2013}.
Similarly, there is a correspondence between the modeling of $k$-atic liquid crystals with vector field design, which has been developed in applications such as mesh and texture generation \cite{sageman-furnas_chebyshev_2019, knoppel_globally_2013, de2016vector}. 
In the discretization we present below, we integrate the finite element approximation with concepts from discrete differential geometry (DDG). 
DDG adopts geometric arguments in constructing geometric invariants, such as curvatures, allowing exact transference of fundamental theorems—such as the Gauss-Bonnet and Poincaré-Hopf theorems—from the smooth geometry to the discrete one.

The rest of the paper is structured as follows. We first introduce the geometric formulation of active nematics in a continuous setting in Sec.~\ref{sec:theory}. 
This is followed by the development of a discretization approach that adheres to the underlying geometric principles in Sec.~\ref{sec:methods}.
Results are discussed in Sec.~\ref{sec:results} , where we verify our numerical scheme through convergence tests and demonstrate its computational efficiency, and where we present numerical solutions for the dynamics of active nematics on diverse geometries, highlighting the ability of our framework to capture the intricate dynamics under various flow regimes and surface topologies. We conclude and discuss potential extensions of this work in Sec.~\ref{sec:conclusion}.

\section{Theoretical formulation \label{sec:theory}}
An active fluid is a soft material whose constituent elements have the ability to exert internal mechanical stresses. 
In systems consisting of microscopic particles, like those found in many biological systems, inertia plays a negligible role due to their small size and low velocities, resulting in very small Reynolds numbers. 
In such cases, the Navier-Stokes equations governing the hydrodynamics simplify to the Stokes equations. 
Here, the fluid flow is driven by the active stresses induced by the constituent elements and is dominated by viscous effects.
To study the collective behavior of these active fluids on large length scales, we use a continuum description of the particles through an order parameter field. 
For example, a polar element can be represented as a vector, denoted as $\bq$.
In active nematics, microscopic constituents exhibit nematic symmetry, meaning there is an equivalence relation between orientations $\bq$ and $-\bq$. 
In 2D, this nematic order, denoted as $[\bq]$, can be equivalently represented by a special real $2 \times 2$ matrix or by a complex-valued function \cite{Pismen1999VorticesIN}, as we explain in more detail below.
The configuration of the nematic field is governed by the nematodynamics equation, which resembles an advection-diffusion equation acting on $[\bq]$. 
The diffusion part of nematodynamics stems from the alignment with neighboring nematic elements through local steric interactions. 
When the nematic particles are suspended in a fluid, they are also advected by the fluid flow and corotate due to fluid deformations.
The time scale separation between the nematodynamics and fluid viscous dissipation allows for an effective model based on a quasi-static coupling between nematodynamics and the Stokes equations.
When a significant amount of energy is inputted by the active components, these systems can exhibit a wide range of phases, ranging from well-ordered patterns to chaotic states.
Because the system operates on a curved Riemannian surface and our discretization method is devoid of global coordinates, we first introduce the system of equations through an abstract geometric language and then present the coordinate-based expression in a later section.

\subsection{Mathematical framework}\label{sec: math framework}
In this section, we will establish the geometric framework required to formulate the hydrodynamics of active nematics on a 2D Riemannian manifold.
In Secs.~\ref{sec: order parameter} and \ref{sec: director field}, we introduce the complex representation of a $k$-atic director, specifically for the tangent vector and the nematic director.
In Sec.~\ref{sec: lie derivative}, we introduce the Lie derivative on tensor fields, which is crucial for describing transport phenomena and deformations in fluid mechanics. 
The Bochner Laplacian governs the relaxation dynamics of vector and nematic fields. 
We start by introducing the Levi-Civita connection (\ie covariant derivative) on the tangent bundle in Sec.~\ref{sec: continuous connection}, which enables us to extend the discussion to the nematic connection in Sec.~\ref{sec:Levi-Civita connection on the nematic bundle} and the Bochner Laplacian on each bundle in Section~\ref{sec: laplacian and diffusion}.
In Sec.~\ref{sec: holomorphic structure}, through the perspective of complex manifolds, we decompose the covariant derivative into Dolbeault operators and establish their connections with fluid kinematics.
Local features like curvature and nematic singularities have an influence on active nematodynamics on a manifold and are related to surface topology through local-to-global theorems in differential geometry.
In Sec.~\ref{sec: continuous topology}, we detect and quantify charges associated with nematic topological defects by measuring the total rotation of a director field along a closed loop using the Levi-Civita connection. 
The curvature emerges as the angle holonomy of this path integration. 

\subsubsection{Order parameter of $k$-atic director}\label{sec: order parameter}
We first introduce the general description for a \(k\)-atic director in the plane, such as a vector ($k = 1$) and a nematic director ($k = 2$).
Let $V$ be a 2D Euclidean vector space, and the multiplicative group action of complex numbers, $\Complex^\times$, on $V$ is given by:
\begin{align}
    \Complex^\times \times V \rightarrow V, \quad (re^{\Img \theta}, \bu) \mapsto 
    \mathrm{Scale}{(\mathrm{Rotate}{(\bu, \theta)}, r)}.
\end{align}
A finite subgroup of rotation actions, $N = \{1, e^{\Img {2 \pi}/{k}}, \ldots, e^{\Img (k -1){2 \pi}/{k}} \} \subset \Complex^\times$, defines an equivalence relation of $k$-rotational symmetry denoted by $\sim$. For any $\bu, \bv \in V$, we have:
\begin{align}
    \bu \sim \bv \quad \mathrm{iff} \quad \exists \quad \coeff z \in N \quad  \mathrm{s.t.} \quad \bv = \coeff z \bu.
\end{align}
This equivalence relation allows us to define a quotient set $V_k = V / N$ and the map of taking the equivalence class $[\cdot]_k: V \rightarrow V_k$.
Each element \([\bq]_k\in V_k\) is called a \(k\)-atic director.
The action of $\Complex^\times$ on $V$ naturally induces a canonical quotient action $\Complex^\times_k = \Complex^\times / N$ on $V_k$, given by:
\begin{align}
    \Complex_k \times V_k \rightarrow V_k, \quad (r^k e^{\Img k \theta}, [\bu]_k) \mapsto [r e^{i\theta} \bu]_k,
\end{align}
where the rotational group is partitioned in $\theta$, with $e^{\Img k \theta} \sim \coeff z e^{\Img k \theta}, \coeff z \in N$.
Formally, we have the power map \( P: \Complex^\times\to\Complex^\times\), \(re^{\Img\theta}\mapsto r^k e^{\Img k\theta}\), which is a homomorphism with kernel being exactly $N$.
By the first isomorphism theorem, $\Complex^\times / \ker(P) \cong \mathrm{im}(P)$, the quotient group is isomorphic to the image of the map, \(\Complex^\times_k \cong\mathrm{im}(P) = \Complex^\times \). 
If given a unit basis vector $\Base \in V$ with $\vert \Base \vert = 1$, we can establish an isomorphism between complex numbers \(\Complex\ni\hat q\) and vectors \(V\ni\bq\) by $\bq = \coeff q \Base = r e^{\Img \theta} \Base \in V$.  
Similarly, given a $k$-director basis $[\Base]_k \in V_k$, we can establish the isomorphism between \(\Complex\ni\hat {[q]}_k\) and its associated $k$-director $ V_k \ni [\bq]_k = \coeff{[q]}_k [\Base]_k = r^k e^{\Img k \theta} [\Base]_k =  \coeff q^k [\Base]_k $.
Through the chain rule, the \emph{pushforward} of the equivalence map, denoted as $d[]_k\vert_{\bq} $, maps an increment vector, $\bq' \in T_{\bq} V $, to  an increment $k$-atic director, $ [\bq]'_k = k  [\bq]_{k-1} \bq' \in T_{[\bq]_k}V_k$.

When focusing on a nematic field with $k = 2$ and $N = \{ 1, -1 \}$, an alternative representation of the equivalence class is through a rank-$1$ symmetric tensor $[\bq]_2 \cong \bq \otimes \bq \in \mathcal{Q}$, where $ \mathcal{Q} = \{  \, \bQ \in V \otimes_{\symm} V \mid \text{rank }\bQ \leq 1  \, \} $.
We called the bijective map from the representation of complex numbers to the matrix representation the \textit{Veronese map}, 
\begin{align}\label{eqn: veronese}
\mathcal{V}: V_2 \rightarrow \mathcal{Q}, 
\quad  [\bq]_2 \mapsto \bq \otimes \bq.
\end{align}
For clarity, in the reminder of this paper, we will omit the subscript and specify the bracket to denote a nematic equivalence (\ie $[] = []_2$). 
With $[\bq] \in V_2$, $\bQ = \mathcal V ([\bq]) \in \mathcal Q$, the pushforward of the Veronese map $d \mathcal{V}:T_{[\bq]} V_2 \rightarrow T_{\bQ} \mathcal Q $ maps an increment $[\bq]'$ in complex representation to the increment in matrix representation $ \bQ'$.

\subsubsection{Vector field and nematic field on a Riemannian manifold}\label{sec: director field}
On a closed Riemannian 2-manifold $M$, $\partial M = \varnothing$, each point $p\in M$ has a tangent space $T_pM$ that is a copy of the Euclidean plane: $T_pM\cong V$.  
Tangent spaces at different points, \(T_pM\), \(T_qM\), for \( p\neq q\), are disjoint spaces, and the disjoint union of all tangent spaces is referred to as the \emph{tangent bundle}, \(TM = \bigcup_p T_pM\).  
Here the tangent space \(T_pM\) is also referred to as the \emph{fiber} of the \emph{fiber bundle} $TM$ at a base point \(p\in M\).
A tangent vector field \(\bq\) is described as a \emph{section} of the tangent bundle, \(\bq\in\Gamma(TM)\), which is an assignment of a fiber element \(\bq_p\in T_pM\) at each point \(p\in M\).
We also denote a section defined over a region \(U\subset M\) as \(\bq\in \Gamma_U(TM)\).

Similarly, we can describe a nematic field as a section of the nematic bundle $L$, $[\bq] \in \Gamma(L)$.
Each fiber of the nematic bundle $L$ at \(p\in M\), $L_p$, is given by the equivalence relation, \(L_p \coloneqq [T_pM] (= [T_pM]_2)\) (c.f.\@ Sec~\ref{sec: order parameter}). 
In other words, the nematic bundle \(L = [TM]\) is obtained by taking fiberwise equivalence class of the tangent bundle.

With a basis section for the tangent bundle, \(\Base\in\Gamma_U(TM), |\Base|=1\), defined in a neighborhood $U\subset M$, we can construct the corresponding nematic basis section for \(L\) as \([\Base] \in \Gamma_U(L)\) .
Basis sections allow us to represent vector fields and nematic fields by complex scalar functions, $\hat q\colon U\to\Complex$ and $\coeff{[q]}\colon U\to \Complex_2$.

Recall that a nematic vector \([\bq]\) can also be represented by a rank-1 symmetric tensor via the Veronese map of Eq.~\eqref{eqn: veronese}. 
We can similarly represent a nematic field as a section of the symmetric rank-1 tensor bundle as follows.
The complex representation of the nematic field is isomorphic to a tensor field $\bQ = \bq \otimes \bq \in \Gamma(E)$, where $E$ is the fiber bundle given by the set \(\mathcal{Q}\) of rank \(\leq 1\) in the symmetric tensor bundle \(TM\otimes_{\rm symm} TM\).
By populating the fiber-wise Veronese map $\mathcal{V}: V_2 \rightarrow \mathcal{Q}$, we get the section-wise map $\mathcal{V}: \Gamma(L) \rightarrow \Gamma(E)$.

The Riemannian metric structure, $\bg|_p: T_pM \times_{\symm} T_pM \rightarrow \Real$, allows a positive-definite inner product $\langle \cdot, \cdot \rangle = \bg(\cdot, \cdot)$ between tangent vector fields or nematic fields.
For any two vector fields $\bu = \coeff u \Base$, $\bv = \coeff v \Base$ and two nematic fields $[\bq] = \coeff{[q]} [\Base] \cong \bq \otimes \bq$, $[\bp] = \coeff{[p]} [\Base] \cong \bp \otimes \bp$, the inner products are defined using their complex representations, 
\begin{align}
    \langle \cdot , \cdot \rangle: 
    \begin{cases}
        V \times V \rightarrow \Real, &\langle \bu, \bv \rangle \mapsto \Re(\coeff u  \overline{\coeff v})\\ 
     V_2 \times V_2 \rightarrow \Real, &\langle [\bq], [\bp] \rangle \mapsto \Re (\coeff{[q]}  \overline{\coeff {[p]}})  \\
     \mathcal{Q} \times \mathcal{Q} \rightarrow \Real, &\langle \bp \otimes \bp, \bq \otimes \bq \rangle = \langle \bp, \bq \rangle^2 \mapsto  \left (\Re ( \coeff q \overline{\coeff p}) \right)^2,
    \end{cases}
\end{align}
where $\overline{\cdot}$ denotes complex conjugation. 
Note that the inner product for $\mathcal{Q}$ is the Frobenius inner product that can be extended and denoted as $\bG= \bg \otimes \bg: (V \otimes V) \times (V \otimes V) \rightarrow \Real$. 
A functional that maps a tangent vector field to a scalar function, $\alpha: T_pM \rightarrow \Real$, is referred to as a \emph{covector} field.
A natural \emph{dual pairing} of $\alpha$ with a vector field $\bu$ is denoted as $\alpha \pair {\bu}$.
We denote this as $\alpha \in \Gamma(T^*M)$, where $T^*M$ is known as the cotangent bundle, which is the dual space to the tangent bundle.
The metric provides a way to associate a vector with its dual counterpart, $\bg|_p: T_pM \rightarrow T_p^*M$, where a vector $\bu$ is mapped to $\bg(\bu, \cdot)$. 
In index notation, this is referred to as \emph{lowering} an index, and it can be expressed as $u_i = g_{ij} u^j$.
Similarly, the Frobenius inner product $\bG|_p: (T_p M \otimes T_p M) \rightarrow (T_p^* M \otimes T_p^*M )$ provides a way to map a $(2, 0)$ tensor to a $(0, 2)$ tensor, or $P_{ij} = g_{im} g_{jn} P^{mn}$ in index notation. 
With metric, we can subsequently denote the $L_2$ inner product of a vector or nematic field on $M$ as $\llangle \cdot, \cdot \rrangle \coloneqq \int_M \langle \cdot, \cdot \rangle ~dA: \Gamma(\circ) \times \Gamma(\circ) \rightarrow \Real$. 
The corresponding metric norm and $L_2$ norm are denoted as $\vert \cdot \vert$ and $\norm{\cdot}$.

\subsubsection{Flow, Lie derivative, and advection}\label{sec: lie derivative}
In differential geometry, the Lie derivative provides a generalized framework for assessing changes in tensor fields encompassing functions with scalar, vector, and covector values, along the flow defined by another vector field \cite{schutz1980geometrical, kanso_geometric_2007}.
In the context of fluid mechanics and in particular complex fluids, the Lie derivative generalizes the concept of directional derivative and henceforth material derivative by capturing the dynamical behavior and transformations of the constituent elements such as the nematic field on a general manifold. 
By employing the Lie derivative, we can establish a comprehensive definition of the advection transport phenomena. 
Notably, when the Lie derivative is applied to a vector field, it recovers the well-known upper-convected derivative. 
Applying the Lie derivative to the metric yields information about the infinitesimal generation of flow velocity deformation. 
This unveils the concept of viscous stress, arising from a fluid's inherent resistance to infinitesimal deviations from isometry.
From a practical standpoint, the Lie derivative perspective provides a unifying geometric discretization that forgoes the use of coordinate-based expressions and respects the underlying continuous structure. 

A velocity field $\bu\in \Gamma(TM)$ generates an instantaneous flowmap $\varphi(t): M \rightarrow M$ such that the time derivative of the flowmap is the velocity field, $\dot{\varphi} = \bu, \varphi (0) = {\rm id}_M$.
The Lie derivative is defined as the rate of change of the so-called pullback map $\varphi^*$ induced by this time-varying flowmap.
Here, we first motivate these concepts by applying the Lie derivative on a scalar function.
A pullback map for a scalar function pulls back a smooth function defined on image $\varphi(M)$, $f \in C^\infty (\varphi(M))$, to a function defined on the domain $M$, $\varphi^* f \in C^\infty(M)$, by precomposition, $\varphi^* f = f \circ \varphi: M \rightarrow \varphi(M) \rightarrow \Real$.
As a result, the Lie derivative on a scalar function leads to the familiar material derivative commonly seen in continuum mechanics,
\begin{align}
\left . \frac{\partial}{\partial t} \right |_{t = 0} \varphi^*f \eqqcolon \frac{\partial f}{\partial t} +  \Lie_{\dot \varphi} f  = \lim_{t \rightarrow 0} \frac{f_{\varphi (p)} - f_p}{t} = \dot f + d_{\bu} f,
\end{align}
where $d_{\bu} f$ is the scalar directional derivative. 
The directional derivative is an example of a natural {dual pairing}, $d_{\bu} f = df \pair{\bu} \in \Real$, between the function map $df|_p: T_pM \rightarrow T_{f(p)}\Real \cong \Real$, with a tangent vector $\bu \in T_pM$ without invoking the metric. 

To define pullback for a general tensor field, we first introduce the pushforward map or tangent map. 
The  flowmap $\varphi$ induces a fiber-wise pushforward map on the tangent space, $ d \varphi|_p: T_{p} M \rightarrow T_{\varphi(p)} M$.
In differential geometry, the pushforward of a map sends a tangent vector $\bq$ in the domain to a tangent vector at the image.  
In the context of continuum mechanics, the pushforward map is referred to as the deformation gradient, where it maps a material fiber in the Lagrangian coordinate to a fiber in the Eulerian coordinates. 
We therefore conveniently denote the vector at $p$ deformed by the flowmap as $d \varphi|_p(\bq) = \bF \bq$.
Populating to the fiber bundle, the pushforward map maps a tangent vector field to another, $d \varphi: \Gamma(TM) \rightarrow \Gamma(TM)$.
Because of the linearity of the tangent map, $d \varphi(-\bq) = -d \varphi(\bq)$, we can deduce that the deformation of the nematic field is equivalent to first deforming its associated vector field and taking the nematic equivalence, or $d \varphi([\bq]) = [d \varphi(\bq)] = [\bF \bq ]$.
This is equivalent to the map on the tensorial representation of the nematic field, $d \varphi(\bq \otimes \bq) = d \varphi (\bq) \otimes d \varphi (\bq) = \bF \bQ \bF^\top$.
Therefore we have the overloaded pushforward map defined over the vector field and both representations of the nematic field, 
\begin{align}
       d \varphi:
    \begin{cases}
      \Gamma(TM) \rightarrow \Gamma(TM), & d \varphi(\bq) = \bF \bq \\
      \Gamma(L) \rightarrow \Gamma(L), & d \varphi([\bq]) =  [\bF \bq] \\
      \Gamma(E) \rightarrow \Gamma(E), & d \varphi(\bQ) = \bF \bQ \bF^\top. 
    \end{cases}
\end{align}
 The pullback map for a covector field pulls back a covector field on $\varphi(M)$ to $M$, $\varphi^*: T^*_{\varphi(p)}M \rightarrow T^*_{p} M$, which is the adjoint operator of the pushforward map.
Therefore for covector field $\Balpha \in \Gamma(T^*M)$, the pullback map is $\varphi^* \Balpha \mapsto (d \varphi)^\top \Balpha =  \bF^{\top} \Balpha$.
Because the metric $\bg(\cdot, \cdot )  \in \Gamma(T^*M \otimes_{\symm} T^*M)$ is a valence $(0, 2)$ tensor, its pullback map follows the covector transformation rule and distributes over the tensor product, or $\varphi^* \bg \mapsto \bF^\top 
\bg \bF$. 
To satisfy the homomorphism property of the pullback map on preserving the dual pairing, $\varphi^*(\alpha \pair{\bq}) = \varphi^* \alpha \pair{\varphi^* \bq}$, a pullback map for a vector field is the inverse map of the pushforward $\varphi^*|_p= (d \varphi|_p)^{-1} = \bF^{-1}: T_{\varphi(p)}M \rightarrow  T_{p}M$.
This fiberwise pullback map pulls back a tangent vector in the Eulerian coordinate to a tangent vector in the Lagrangian coordinate.
Similar to the pushforward map, we can populate the pullback map to sections of bundles.
With omission of the precomposition for non-scalar fields (\eg \({\bq}\circ\varphi\) is denoted as \(\bq\)), we can denote pullback maps as 
\begin{align}\label{eqn: continuous pullback}
       \varphi^*:
    \begin{cases}
        C^{\infty}(M) \rightarrow C^{\infty}(M), & \varphi^*f = f \circ \varphi\\
        \Gamma(G) \rightarrow \Gamma(G), & \varphi^* \bg = \bF^\top \bg \bF \\ 
      \Gamma(TM) \rightarrow \Gamma(TM), & \varphi^*\bq = \bF^{-1} \bq \\
      \Gamma(L) \rightarrow \Gamma(L), & \varphi^*[\bq] = [\bF^{-1} \bq] \\
      \Gamma(E) \rightarrow \Gamma(E), & \varphi^*\bQ = \bF^{-1} \bQ \bF^{-\top}. 
    \end{cases}
\end{align}
The Lie derivative measures the rate of change of this time-varying pullback field at time equals zero,
\begin{align}\label{eqn: Lie derivative as derivative of pullback}
\left.\frac{\partial}{\partial t}\right|_{t=0}\circ \varphi^* \eqqcolon \frac{\partial}{\partial t} +  \Lie_{\dot \varphi}:   
\begin{cases}
        \dot f + \Lie_{\bu} f &= \lim_{t \rightarrow 0} \left ( f_{\varphi (p)} - f_p \right) / t \\
       \Lie_{\bu} \bg &= \lim_{t \rightarrow 0} \left ({\bF^{\top}\bg_ {\varphi (p)} \bF - \bg_p } \right) / t  \\
      \dot \bq + \Lie_{\bu} \bq &= \lim_{t \rightarrow 0} \left ({\bF^{-1}  \bq_ {\varphi (p)} - \bq_p } \right) / t  \\
       \dot {[\bq]} + \Lie_{\bu} [\bq] &= \lim_{t \rightarrow 0} \left ({ \left[\bF^{-1}  \bq_{\varphi (p)} \right] - [\bq]_p } \right) / t  \\
       \dot {\bQ} + \Lie_{\bu} \bQ &= \lim_{t \rightarrow 0} \left ({ \bF^{-1}  \bQ_{\varphi (p)} \bF^{-\top} - \bQ_p } \right) / t.  \\
\end{cases}
\end{align}
Here, we assume that the metric is time-independent, $\dot{\bg} = 0$.
The Lie derivative of the metric, $\Lie_{\bu} \bg$, is commonly known as the strain rate tensor in continuum mechanics. 
It represents the rate of change of the finite strain characterized by the Green-Lagrange strain tensor, $\bF^\top \bg \bF - \bg$.
It is important to note that the rate of change of a pullback field generates an increment-valued field. 
For instance, $\Lie_{\bu} \bq$ is a $T_{\bq} V$-valued vector field. 
 Consequently, the Lie derivative does not commute with nematic equivalence, that is, $\Lie_{\bu} [\bq] \neq [\Lie_{\bu} \bq]$.
Similarly, $\Lie_{\bu} \bQ$ is a tensor field valued in $T_{\bq} \mathcal{Q}$, and it preserves the rank-1 structure of the tensorial representation of nematics. 
When the pullback field is constant, or ${\partial \varphi^*}/{\partial t} = 0$, the section is referred to as being \emph{Lie advected} by the flow, $\dot{\varphi} = \bu(t)$. In the scalar case, this recovers the familiar material derivative.
When the Lie derivative is applied to a vector and nematic field, the deformation caused by the velocity gradient of the flow will rotate and stretch the director fiber. 
This phenomenon is commonly referred to as the upper-convected derivative or Oldroyd derivative.  

In summary, the Lie derivative serves as a generalization of the concept of advection from scalar functions to tensor fields. 
It allows us to extend the idea of a directional derivative, originally defined for scalar functions, to general tensor fields within a continuous deformation generated by a flow represented by a vector field. 
This Lie derivative of the director field remains well-defined even in the absence of the Riemannian metric.  
However, it is crucial to note that, unlike directional derivatives, the Lie derivative with respect to a \emph{single} vector is ill-defined due to the lack of a well-defined deformation map.
To extend the concept of single-vector directional derivative on a manifold from scalar fields to vector or tensor fields, additional structures such as the connection or parallel transport become essential.
The resulting derivative is known as the covariant derivative. 

\subsubsection{Levi-Civita connection on the tangent bundle}\label{sec: continuous connection}
In this section, we cover the necessary background on the Levi-Civita connection similar to other expositions on differential geometry \cite{needham2021visual}.
On a curved space, each tangent space is in general different from other tangent spaces.
The Levi-Civita connection introduces the concept of parallel transport, allowing us to compare vectors at different points along a curved manifold. 
By establishing this connection, we can define the covariant derivative $\cvr$ that extends the concept of differentiation of a tensor field from flat spaces to curved spaces.

Again, we first motivate the concept of covariant derivative using a scalar function. 
Given a parametrized segment on $M$, $\gamma(t): [-\epsilon, \epsilon] \rightarrow M$, with the starting point $\gamma(0) = p$ and velocity $\gamma'(0) = \bv$, the directional derivative of a scalar function $f: M \rightarrow \Real$ with the \emph{single} vector $\bv$ is 
\begin{align}
   d_{\bv} f \coloneqq  \left. \frac{d}{dt} f (\gamma(t)) \right \vert_{t = 0}  
    = \lim_{t \rightarrow 0} \frac{f(\gamma(t)) - f(p)}{t }.
\end{align}

To extend the concept of directional derivatives, initially applicable to scalar functions, to functions represented as tangent vectors $\bu \in \Gamma(TM)$, we must address the challenge of comparing vectors in different tangent spaces: $\bu(\gamma(t))$ in $T_{\gamma(t)} M$ and $\bu(p)$ in $T_{p}M$, where generally $T_{\gamma(t)} M \neq T_{p}M$.
To overcome this challenge, we employ the concept of parallel transport, which is a one-parameter family of linear maps dependent on the path $\gamma$, denoted as $\Pi_{p\rightsquigarrow\gamma(t)}\colon T_{p}M\to T_{\gamma(t)}M$, with $t \in [-\epsilon,\epsilon]$.
For a 2-manifold $M$ equipped with a Riemannian metric, we can intrinsically characterize parallel transport. Specifically, when transporting along a geodesic segment $\gamma$, $\Pi_{p\rightsquigarrow\gamma(t)}\bu(p)$ maintains its length and forms a constant angle with respect to the tangent vector $\gamma'(t)$. In cases where $\gamma$ is not a geodesic, the rate of change of the angle between $\Pi_{p\rightsquigarrow\gamma(t)}\bu(p)$ and $\gamma'(t)$ is determined by the negative geodesic curvature of $\gamma$ at $\gamma(t)$.
Using this notion of parallel transport, we define the directional covariant derivative of $\bu$ with respect to $\bv$ as follows:
\begin{align}
\cvr_{\bv} \bu = \left. \frac{\cvr}{dt} \bu (\gamma(t)) \right \vert_{t=0}
\coloneqq \lim_{t\rightarrow 0} \frac{\Pi_{\gamma(t)\rightsquigarrow p} \bu(\gamma(t)) - \bu(p)}{t},
\end{align}
where $\Pi_{\gamma(\epsilon)\rightsquigarrow p}$ denotes the inverse of $\Pi_{p\rightsquigarrow \gamma(\epsilon)}$.
Note that the covariant derivative is uniquely defined when $\bv = \gamma'(0)$ and is independent of the choice of $\gamma$ as long as $\gamma(0) = p$ and $\gamma'(0) = \bv$. The use of $\cvr$ in place of $d$ signifies the incorporation of infinitesimal parallel transport.
The covariant derivative, denoted as $\cvr \bu$, is obtained by removing the directional vector from the directional covariant derivative such that $\cvr \bu \pair{\bv} \coloneqq \cvr_{\bv} \bu$ or, in index notation, as $v^i \cvr_i u^j$. 
The covariant derivative maps a vector field to a vector-valued covector field, represented as $\cvr: \Gamma(TM) \rightarrow \Gamma(TM \otimes T^*M)$.

Alternatively, when considering the surface being isometrically embedded in 3D Euclidean space through an embedding function $\br: M \rightarrow \Real^3$, we can intuitively understand parallel transport via the ambient 3D space.
The embedding function induces a pushforward map, $d \br|_p: T_p M \rightarrow \Real^3$, which realizes each tangent vector, $\bu \in T_p M $, as an Euclidean vector, $d \br|_p \pair{\bu} \in \Real^3$.
The concept of parallel transport along a path $\gamma$ that connects point $p = \gamma(0)$ to $\gamma(\epsilon)$ can be geometically described by rolling the tangent plane at \(\gamma(0)\) along the path \(\gamma\) to \(\gamma(\epsilon)\) against the surface without slipping.
Because this rolling process is typically well-understood in a discrete setting, it is applied to discretize the covariant derivative in Sec.~\ref{sec: gradient and covariant divergence}.

Recall that under a local basis section, $\Base\in\Gamma_U(TM)$, where $|\Base|=1$, a tangent vector field $\bu\in\Gamma_U(TM)$ can be represented as a complex field $\coeff u\colon U\to\Complex$, given by $\bu = \coeff u\Base$. In this complex representation, the covariant derivative can be expressed as
\begin{align}\label{eqn: covariant derivative}
    \cvr  \bu = (d  \coeff u) \Base + \coeff u (\cvr  \Base) \eqqcolon \left((d+\Img\omega)\hat u \right)\Base.
\end{align}
where $d \coeff u = d  \Re{(\coeff u)} + \Img d  {\Im{(\coeff u)}}  \in \Gamma_U(T^*M; \Complex)$ represents a complex-valued covector field and \(\omega\) is a real-valued covector field \(\omega\in\Gamma_U(T^*M; \Real)\) defined by the covariant derivative of the basis field \(\cvr\Base \eqqcolon \Img \omega\Base \in \Gamma_U(TM \otimes T^*M)\).
The components of $\omega$ are commonly referred to as the Christoffel symbol.

\subsubsection{Holomorphic structure with Levi-Civita connection \label{sec: holomorphic structure}}

As being demonstrated repeatedly, a vector field, when equipped with a basis field, can be viewed as a complex scalar field.
Using the Levi-Civita connection as the spatial derivative operator for this complex field, we can distinguish its holomorphic and anti-holomorphic components by orthogonally decomposing the covariant derivative into the \emph{Del} and \emph{Del bar} operators, $\nabla = \partial^{\cvr} + \bar \partial^{\cvr}$, 
\begin{align}
\begin{split}
      \bar{\partial}_{\bv}^{\cvr} \bu &\coloneqq \frac{\cvr_{\bv} \bu + \Img \cvr_{\Img \bv} \bu}{2}  \\
      \partial_{\bv}^{\cvr} \bu &\coloneqq \frac{\cvr_{\bv} \bu - \Img \cvr_{\Img \bv} \bu}{2}.
\end{split}
\end{align}
Note that $\Img$ represents rotation of the vector by $90^\circ$, $\Img(\cdot) = \operatorname{Rotate}(\cdot,90^\circ)$.
A vector field \(\bu\) is holomorphic if and only if \(\nabla\bu\llbracket\Img \cdot\rrbracket = \Img\nabla\bu\llbracket\cdot\rrbracket\), or $\bar \partial^{\cvr} \bu = 0 $. 
Its conjugate $\partial^{\cvr} \bu = 0$ indicates $\bu$ is antiholomorphic.

It turns out that the holomorphic and anti-holomorphic parts play imprtant roles when the vector field $\bu$ represents the velocity field of a surface flow.
When $\bu$ denotes the fluid velocity, the holomorphic part, $\partial^{\cvr} \bu$, quantifies the extent of conformal deformation (involving infinitesimal rotation with local dilation) of the fluid.
The anti-holomorphic part, $\bar \partial^{\cvr} \bu$, corresponds to the symmetic and trace-free part of $\cvr \bu$.
It is the strain rate tensor for incompressible flow and measures the infinitesimal shearing that viscosity resists.
This concept will be applied in Sec.~\ref{sec: forced stokes flow} when formulating the Stokes equations on the manifold.

\subsubsection{Levi-Civita connection on the nematic bundle} \label{sec:Levi-Civita connection on the nematic bundle}
We can extend the Levi-Civita connection from the tangent bundle to the nematic bundle.
This is equivalent to defining a nematic covariant derivative, denoted by $\cvr^L: \Gamma(L) \rightarrow \Gamma(L \otimes T^*M)$.
Under a local nematic basis $[\Base] \in \Gamma_U (L)$, the nematic covariant derivative can be defined as:
\begin{align}\label{eqn: nematic covariant derivative}
\cvr^L [\bq] = d \coeff{[q]} \Base + \coeff{[q]} \cvr^L [\Base] \coloneqq \left ((d+\Img 2\omega)\coeff{[q]} \right)[\Base],
\end{align}
 where $\omega$ is the covector field $\Img\omega\Base\coloneqq\nabla\Base$ associated to the tangent bundle connection and basis. 
 The nematic connection $\cvr^{L} [\Base] \coloneqq \Img 2 \omega [\Base]$ reflects increased rotational speed due to nematic symmetry. 
 This definition is basis-independent and aligns with the algebraic construction of the Levi-Civita connection on the tangent bundle, $\nabla^{L} [\bq] = 2 \bq \nabla \bq = \nabla (\bq^2)$.

The nematic covariant derivative $\nabla^L[\bq]$ is consistent with the standard covariant derivative for matrix-valued functions, $\cvr: \Gamma(TM \otimes TM) \rightarrow \Gamma(TM \otimes TM \otimes T^*M)$, when applied to $ \bQ = \mathcal{V} ([\bq]) \in \Gamma(E) \subset \Gamma(TM \otimes TM)$. 
Like the Lie derivative in the advection equation (c.f. discussion on Eqn.~\eqref{eqn: Lie derivative as derivative of pullback}), covariant derivatives also produce increment-valued nematic fields, $\nabla^L[\bq]$ and $\nabla \bQ $.
Via the pushforward of the Veronase map, $d \mathcal{V}|_{{[\bq]}}: T_{{[\bq]}} \Gamma(L) \rightarrow T_{\mathcal{V}({[\bq]})} \Gamma(E) $, one can confirm the equivalence of these two derivatives, $\nabla \bQ = \nabla \bq \otimes \bq + \bq \otimes \cvr \bq = d\mathcal{V}|_{[\bq]} \pair {\nabla^L [\bq]}$.

\subsubsection{Laplacian and diffusion}\label{sec: laplacian and diffusion}
The concept of the covariant derivative naturally leads to a Laplacian operator referred to as the Bochner Laplacian, which is the first variation of the Dirichlet energy.

To begin, we introduce the Bochner Laplacian when it operates on vector fields. 
In the context of liquid crystals in Euclidean space, the Dirichlet energy of a vector field is also known as the one constant Oseen-Frank free energy of a polar liquid crystal \cite{de1993physics, frank1958liquid}.
The Dirichlet energy functional maps a vector field to an energy measure, $\mathcal F: \Gamma(TM) \rightarrow \Real$, and can be expressed as follows:
\begin{equation}\label{eqn: dirichlet}
\mathcal{F}(\bu) 
=  \frac{1}{2} \norm{ \cvr \bu}^2 
\overset{\partial M = \varnothing}{=}  \frac{1}{2} \llangle \bu , \cvr^* \cvr \bu  \rrangle  \eqqcolon -\frac{1}{2} \llangle \bu , \blp \bu  \rrangle,
\end{equation}
where the operator $\cvr^*: \Gamma (TM \otimes T^*M) \rightarrow \Gamma(TM)$ represents the $L_2$ adjoint of $\cvr$, defining the Bochner Laplacian as $\blp \coloneqq - \nabla^* \cvr: \Gamma (TM) \rightarrow \Gamma(TM)$.
Note that we use the negative semi-definite convention for the Laplacian operator.

Using the same procedure, the nematic covariant derivative is employed to construct the nematic Dirichlet energy, $\mathcal{F}^L \in \Gamma(L) \rightarrow \Real$, which is expressed as:
\begin{align}
 \mathcal{F}^L([\bq]) 
 = \frac{1}{2} \norm {\cvr^L [\bq]}^2
 \overset{\partial M = \varnothing}{=}  \frac{1}{2} \llangle {[\bq]} , \cvr^{L*}\cvr^L [\bq]  \rrangle  \eqqcolon -\frac{1}{2}  \llangle {[\bq]} , \blp^L [\bq]  \rrangle.
\end{align}
The variation of $\mathcal F^L$ leads to the definition of the nematic Bochner Laplacian, denoted as $\blp^L \coloneqq -\cvr^{L*}\cvr^L: \Gamma(L) \rightarrow \Gamma(L)$.

Analogously, the standard Bochner Laplacian $\Delta: \Gamma(TM \otimes TM) \rightarrow \Gamma(TM \otimes TM)$ defined for a general matrix field $\bH$ can be derived from the matrix Dirichlet energy, $\mathcal{E}: \Gamma(TM \otimes TM) \rightarrow \Real$, measured by the Frobenius inner product, $\mathcal{E}(\bH) = \frac{1}{2} \norm {\nabla \bH}^2 =  \frac{1}{2} \int_{M} \nabla_i H_{jk} \nabla ^i H^{jk} dA$.
This Laplacian $\Delta$ can act on a nematic field $\bQ = \mathcal{V}([\bq])$, but it differs from the nematic Bochner Laplacian, $\Delta \bQ \neq d\mathcal{V}|_{[\bq]}(\Delta^L [\bq])$.

The reason for this difference is two-fold. 
First, nematic fields only correspond to rank-1 matrices in the nematic subspace $E \subset  TM\otimes TM$, but the matrix Bochner Laplacian spans the entire \(TM\otimes TM\). This results in the matrix representation of nematic fields being an overparametrization under this operator.
This overparametrization can be demonstrated by considering the matrix diffusion equation $\dot \bQ =\blp \bQ$.
In general, a rank-1 initial condition $\bQ|_{t=0} \in \Gamma(E)$ does not remain rank-1 under this diffusion (\ie $\bQ$ diverges from the nematic subspace $E$). 
In contrast, $\dot{[\bq]} = \blp^L[\bq]$ remains within the nematic phase. 
Second, despite the equivalence of covariant derivatives $\nabla \bQ = d \mathcal{V}|_{[\bq]} \pair{ \nabla^L [\bq] }$ (\cf Sec.~\ref{sec:Levi-Civita connection on the nematic bundle}), their corresponding Dirichlet energies are distinct: $ \mathcal{E}(\bQ) \neq \mathcal{F}^L([\bq])$. 
This is because the pushforward of the Veronese map \(d\mathcal{V}_{[\bq]}\colon T_{[\bq]}L\to T_{\bQ}E\) does not isometrically map the intrinsic norm on $T_{[\bq]}L$ to the Frobenius norm for matrices on $T_{\bQ}E$.

As a side remark, it is possible to represent the nematic Dirichlet energy \(\mathcal{F}^L([\bq])\) in terms of $\bQ$ through an equivalent functional $\mathcal{F}^E\colon \Gamma(E)\to\Real$ such that $\mathcal{F}^E(\bQ) = \mathcal F^L([\bq])$. 
We can achieve this by introducing a specific quadratic form, $ |\cdot|_{\mathcal Q}^2\coloneqq |\cdot|^2 - 2 \det(\cdot)$, for $\Real^{2 \times 2}$ matrices on $TM \otimes TM$.
This quadratic form induces a corresponding symmetric bilinear form $\langle\cdot,\cdot\rangle_{\mathcal{Q}} = \langle \cdot, \cdot \rangle - 2 \det(\cdot, \cdot)$.
The equivalent nematic Dirichlet energy $\mathcal{F}^E$ relies on this bilinear form and can be explicitly expressed in index notation:
\begin{align}\label{eqn: nematic Q dirichlet energy}
    \mathcal{F}^E(\bQ) = \frac{1}{2}\int_M | \nabla \bQ |_{\mathcal Q}^2 ~dA
    \equiv  {\frac{1}{2}}\int_M g^{ij}(g_{km}g_{\ell n} -\det(\bg)\epsilon_{km}\epsilon_{\ell n})\nabla_i Q^{k\ell}\nabla_j Q^{mn}\, dA.
\end{align}
Here, $g_{ij}$ are the components of the metric tensor $\bg$ and $\epsilon_{ij}$ represents the Levi-Civita permutation symbol.

In this paper, we adopt the more natural choices of $\mathcal{F}^L([\bq])$ and $\blp^L$ to model the nematic field. 
We use \(\blp^E\colon\Gamma(E)\to\Gamma(E)\) to denote the Laplacian equivalent to \(\blp^L\) but acting on the matrix field representation.
That is, $\blp^E$ satisfies the equivalence $\blp^E \bQ = \blp^E(\mathcal{V}[\bq]) = d\mathcal{V}\pair {\blp^L[\bq]} \in T_{\bQ} \Gamma(E)$ and is defined as,
\begin{align}\label{eqn: Laplacian E}
  \blp^E \bQ \coloneqq - \frac{\delta \mathcal{F}^E({\bQ})}{ \delta {\bQ|_{E}}} 
  =  -d \mathcal{V} \left \llbracket \frac{\delta \mathcal{F}({[\bq]})}{ \delta {[\bq]}} \right \rrbracket 
 =  {d \mathcal{V} \circ  \blp^L  \circ \mathcal{V}^{-1}} (\bQ).
\end{align}
The subscript $|_E$ emphasizes the variation within the nematic subspace $E$, rather than the entire $TM \otimes TM$.

\subsubsection{Curvature, defects, and local-global geometry}\label{sec: continuous topology}
The singularities within the nematic field, also referred to as topological defects, are associated with points where the director field exhibits zeros and discontinuities. 
These defects are commonly characterized by their index or charge $\mathcal{Z}$.
To identify these defects, we employ the covariant derivative in closed-loop integration, a classic procedure also discussed in other texts \cite{needham2021visual, needham2023visual}.
Interestingly, poles and zeros of complex functions can be naturally viewed as the singularies in director fields.
Power functions, $f(z) = {(z - z_0)}^\mathcal{Z}$, where $f:\Complex \rightarrow \Complex$, are an archetypal demonstration of defects with a charge $\mathcal{Z}$ at $z = z_0$, with $\mathcal{Z}$ being an integer or a fraction.  
On a manifold, the Poincaré-Hopf theorem asserts that the total charge of defects is determined by a topological invariant called the Euler characteristic of the surface, thus establishing a fundamental connection between the nematic field and the topology of the underlying surface.

The charge of a topological defect in a $k$-atic field $[\bq]_k$ is determined by the cumulative turning angle that the director field undergoes around the defect. 
Here we consider a defect at point $p \in U$, where $U\subset M$ represents a sufficiently small disk that does not encircle any other singularities.
The boundary of the disk is denoted as $\gamma = \partial U$.
Locally in $U$ it is always possible to establish a smooth and defect-free basis $[\Base]_k$ and subsequently express the rotation form under the Levi-Civita connection, denoted as $\eta \in \Gamma(T^*M)$, as
\begin{align}\label{eqn: rotation form}
    \eta
    \coloneqq \frac{1}{|[\bq]_k|^2} \left \langle \cvr [\bq]_k, \Img [\bq]_k \right \rangle 
    = \frac{1}{|[\bq]_k|^2} \left \langle d [\bq]_k + \Img k \omega [\bq]_k, \Img [\bq]_k \right \rangle 
    = d^{\mathbb{S}^1} \arg(\coeff{[q]}_k) + k \omega.
\end{align}
Note that the function $\arg$ necessitates a smooth basis to be well-behaved and returns values in $\mathbb{S}^1$ rather than real numbers. 
Despite being \(\mathbb{S}^1\)-valued by itself, the differential, $d^{\mathbb S^1} \arg(\coeff{[q]}_k) $, can be defined as a real-valued function. 
This is because, at a given point, this differential depends only on the function's values in a small neighborhood around the point.
In such a local context, a continuous \(\mathbb{S}^1\)-valued function can be unwrapped to a real-valued function up to a \(2\pi\)-integer constant that does not affect the differential. 

The closed-loop integral of the turning rate $\eta$ defined in Eqn.~\eqref{eqn: rotation form} along the parametrized boundary loop $\gamma(t) = \partial U(t)$, where $t \in [0, T]$ and $\gamma(0) = \gamma(T)$ is referred to as the cumulative turning angle of the director field around a defect.
The integral of the connection $\omega$ in Eqn.~\eqref{eqn: rotation form}, which is known as the holonomy angle, is equal to the negative of the integrated Gaussian curvature: $\oint_{\gamma} \omega  = -\int_U K dA$.
Therefore, the charge of the topological defect in a $k$-atic director field, denoted as $\mathcal{Z} \in \mathbb{Z}/k$, admits the following relation:
\begin{align}\label{eqn: charge}
     \mathcal{Z}_p([\bq]_k) \coloneqq \frac{1}{ 2 \pi k}\oint_{\gamma} d^{\mathbb{S}^1} \arg(\coeff {[q]}_k) = \frac{1}{ 2 \pi k}\oint_{\gamma} \eta  + \frac{1}{2 \pi}\int_U K dA.
\end{align}
Note that the closed integral ensures that the orientation after integration, $\arg(\coeff {[q]}_k)|_{t = T}$, differs only by an integer multiple of $2\pi$ from its starting value, $\arg(\coeff {[q]}_k)|_{t = 0}$.
As a result, the charge is consistently an integer multiple of $1/k$ (\ie an integer value for a vector field and a multiple of $1/2$ for a nematic field).
This quantized property also implies that the charge of the defect remains invariant regardless of variations in the size and shape of $\gamma$ and is solely the property of the director field.
As a special case, \(\oint_{\gamma} d^{\mathbb{S}^1}\arg(\coeff{[q]}_k) = 0 \) if $\gamma$ does not enclose any topological defects of \(\arg(\coeff{[q]}_k)\).  

Owing to the Poincaré-Hopf Theorem, we can relax the local constraint on $U$.
This theorem relates the sum of charges at each point $p_i$ on the entire manifold $M$ (with $\partial M = \varnothing$) to a fundamental topological invariant known as the Euler characteristic $\chi$:
\begin{align}
    \sum_{i} \mathcal{Z}_{p_i}([\bq]_k) = \chi(M) = 2 - 2g, 
\end{align}
where $g$ represents the genus of the manifold.
For instance, this theorem establishes the absence of a nonvanishing continuous director field on a topological 2-sphere.
By replacing $U$ with manifold $M$, Eqn.~\eqref{eqn: charge} leads to the Gauss-Bonnet Theorem, which states that the total Gaussian curvature is also a topological invariant, $\int_M K dA = 2 \pi \chi(M)$.

In a nutshell, we can detect defects and calculate their charges by establishing the cumulative turning angle through the use of the covariant derivative. 
Local-to-global theorems, including the Gauss-Bonnet and Poincaré-Hopf theorems, connect local properties like curvature of the surface and charges of the nematic field to robust global invariants. 
Practically, as we will elaborate in Sec.~\ref{sec: discrete curvature}, these formulas translate to the discretization of Gaussian curvature and the algorithmic computation of topological charges, which satisfy the discrete analog of local-to-global theorems.

\subsubsection{Expressing the Lie derivative as a covariant derivative}\label{sec: Lie derivative as covariant derivative}
Under the Levi-Civita connection, the Lie derivative can be expressed in terms of the covariant derivative. To demonstrate this, we denote the directional derivative, a derivation associated with its corresponding vector field, as $\bu \coloneqq d_{\bu}: C^\infty(M) \rightarrow C^\infty(M)$. In index notation, it is commonly denoted as $u^i \partial_i$.
The pullback map respects the derivation structure; for $ \bq \in \Gamma(TM)$, $f \in C^\infty(M)$, we have $\varphi^*(d_{\bq} f) = d_{\varphi^* \bq} \varphi^* f$.
Consequently, the Lie derivative satisfies the Leibniz rule over derivation, $\Lie_{\bu}(d_{\bq} f) = d_{\Lie_{\bu}\bq}f + d_{\bq}\Lie_{\bu}f$.
Since the Lie derivative of a scalar field is equivalent to the directional derivative, we have $\Lie_{\bu} \bq (f) \coloneqq d_{\Lie_{\bu} \bq} f  = \Lie_{\bu} (d_{\bq} f) - d_{\bq} \Lie_{\bu} f = \bu \bq (f) - \bq \bu (f) \eqqcolon [\bu, \bq](f) $, known as the Lie bracket. 
The Lie bracket can in fact be expressed using the covariant derivative of the Levi-Civita connection, referred to as the torsion-free condition of the connection, $T^\nabla \pair {\bu, \bq} \coloneqq \cvr_{\bu} \bq - \cvr_{\bq} \bu - [\bu, \bq] = 0$.
Under the Levi-Civita connection, the Lie material derivative for vector $\bq$ can be expressed as
\begin{align}
  \dot \bq + \Lie_{\bu} \bq  = \dot \bq + \cvr_{\bu} \bq - \cvr_{\bq} \bu,
\end{align}
which is a familiar form of the upper-convected derivative, or Oldroyd derivative for vectors \cite{oldroyd1950formulation}.
The term $\cvr_{\bu} \bq$ represents the parallel transport components, while $\cvr_{\bq} \bu$ accounts for the local deformation caused by the differential flow velocity. 

Through the pushforward of $k$-atic equivalence $d[]_{k}$, the Lie derivative of the $k$-atic director field can be expressed as follows:
\begin{align}
 \dot {[\bq] }_k + \Lie_{\bu} [\bq]_k    = k \bq^{k - 1}  ( \dot \bq + \Lie_{\bu} \bq).
\end{align}
When $k=2$, this is equivalent to the advection of the tensorial representation of the nematic field. 
Applying the Leibniz rule of the Lie derivative to the tensor product, we have $\Lie_{\bu} \bQ = \Lie_{\bu} \bq \otimes \bq + \bq \otimes \Lie_{\bu} \bq$. 
By rearranging the terms, we recover the operator for the upper-convected derivative of the symmetric tensor $\bQ$,
\begin{align}
  \begin{split}
    \dot \bQ +  \Lie_{\bu} \bQ  =\dot{\bQ} + \cvr_{\bu} \bq \otimes \bq + \bq \otimes \cvr_{\bu} \bq - \cvr_{\bq} \bu \otimes \bq - \bq \otimes \cvr_{\bq} \bu =  \dot{\bQ} +  \cvr_{\bu} \bQ -  (\cvr \bu) \bQ  - \bQ (\cvr \bu^\top).
\end{split}  
\end{align}
The dual pairing order follows the convention of matrix multiplication.
In index notation, this can be expressed as $(\Lie_{\bu}\bQ)^{ij} = u^k \nabla_k Q^{ij} -\nabla_k u^i  Q^{kj} - Q^{ik} \nabla_k u^j$. 

As demonstrated in Section~\ref{sec: lie derivative}, the strain rate tensor can be formulated as the Lie derivative of the metric, $\Lie_{\bu} \bg$. 
When expressing it in terms of the covariant derivative acting on the Eulerian velocity, we can retrieve its well-known form as the symmetric component of the velocity gradient.
By expressing $\dot \bF = \dot{(d \varphi)} = \varphi^*(\cvr \dot \varphi) = (\cvr \bu) \bF \in \Gamma(TM \otimes T^*M)$, we can express the Lie derivative of the metric as the pullback of the strain rate tensor,
\begin{align}
    \Lie_{\bu} \bg 
    = \frac{\partial}{\partial t} (\bF^\top \bg \bF) = \dot \bF^\top \bg \bF + \bF^\top \bg \dot{\bF} 
    &= \bF^\top (\cvr \bu^\top \bg + \bg \cvr \bu) \bF = \varphi^* (\cvr \bu^\top \bg + \bg \cvr \bu).
\end{align}
Since the flowmap, $\varphi: M \rightarrow M$, is instantaneously an identity map, we can express  $\Lie_{\bu} \bg = \cvr \bu^\top \bg + \bg \cvr \bu \in \Gamma(T^*M \otimes_{\symm} T^*M)$, which measures the infinitesimal deformation rate of a fluid element.
This Lie derivative defines the Killing operator, 
\begin{align}
    \mathcal{K}: \Gamma(TM) \rightarrow \Gamma(T^*M \otimes_{\operatorname{symm}} T^*M),\quad \mathcal{K}\colon \bu \mapsto \frac{1}{2}\Lie_{\bu} \bg  .
\end{align}
A vector field \(\bv\) is referred to as a Killing vector field if \(\mathcal{K}\bv = 0\) (\ie the flow it generates preserves the metric).

\subsection{Hydrodynamics of active nematic films}\label{sec: active nematics}
In this section, we will extend the mathematical framework from the previous sections to describe the hydrodynamics of an active nematic film on a curved surface.
The system can be analyzed in two parts. 
First, we have the steady-state Stokes equations, which describe how the fluid responds to the active stress induced by the nematic configuration.
Subsequently, the steady-state fluid velocity acts as the driving force for the advection-diffusion equation that governs the behavior of the nematic field. 
This equation is commonly referred to as the nematodynamics equation.
The system assumes quasi-static coupling, meaning that the dissipation of the fluid at this length scale is considered instantaneous.

\subsubsection{Forced surface Stokes flow \label{sec: forced stokes flow}}
We assume the active nematic is confined to a fluid interface whose flow is governed by the overdamped Stokes equations.
The theoretical foundation for understanding such dissipation-driven systems can be provided by the Onsager variational principle or the Helmholtz minimum dissipation theorem \cite{doi_soft_2013, doi_onsagers_2011, arroyo_balaguer_onsagers_2018, torres-sanchez_modelling_2019, mirza2023variational, arroyo_relaxation_2009}.
The associated functional, known as the Rayleigh dissipation function, captures the dissipation rate of the system, and Stokes flow emerges as the stationary condition of this dissipation function. 
The fluid flow is driven by the stresses exerted by the active nematic. 
In the present work, we only account for the active stress and neglect passive stresses associated with nematic elasticity.
The active stress aligns with the orientation of the nematic constituents and is represented by the stress tensor $\Bsigma = \alpha \bQ \in \Gamma(TM \otimes_{\symm} TM)$, which can be either contractile ($\alpha > 0$) or extensile ($\alpha < 0$) \cite{saintillan_rheology_2018}.
According to Onsager's variational principle, the formulation of Stokes flow involves minimizing the Rayleigh dissipation potential while satisfying the incompressibility constraint, 
\begin{align}
\min_{\bu} \quad 
 \mu \llangle \mathcal K \bu, \mathcal{K} \bu  \rrangle - \llangle \bu, \div^{\cvr} \Bsigma \rrangle \quad
\text{subject to}
  \quad \div~\bu = 0,
\end{align}
where the first term $\mu \llangle \mathcal{K} \bu, \mathcal{K} \bu \rrangle$ quantifies the rate of viscous dissipation \cite{chan_formulation_2017, samavaki_navierstokes_2020, pearce_geometrical_2019}.
Here the divergence operator, $\mathrm{div} \coloneqq -\mathrm{grad}^*: \Gamma(TM) \rightarrow C^{\infty}$, is the negated $L_2$ adjoint of the gradient operator, $\mathrm{grad} \coloneqq \bg^{-1}d: C^{\infty} \rightarrow \Gamma(TM)$.

The negated adjoint of the covariant derivative is referred to as as the covariant divergence, given by $\text{div}^{\cvr} \coloneqq -\cvr^*: \Gamma(TM  \otimes TM) \rightarrow \Gamma({TM})$.

Note that the Killing operator can be factored out using the relation $\mathcal K \bu = \delbar \bu + \div (\bu) \bI / 2$, where $\div (\bu) \bI / 2$ isolates the trace of the strain rate tensor that characterizes fluid dilation (c.f. Sec~\ref{sec: holomorphic structure}). 
Therefore, we can replace $\mathcal K$ with $\delbar$ under the incompressibility constraint and represent the constrained optimization problem as a minimax problem:
\begin{align}
\begin{split}
       \min_{\bu} \max_{p}~ \mathcal{R}  &=   \mu \llangle \delbar \bu, \delbar \bu  \rrangle - \llangle  \bu, \div^{\cvr} \Bsigma \rrangle - \llangle p, \text{div } \bu \rrangle.
\end{split}
\end{align}
In this expression, $p$ is the fluid pressure acting as a Lagrange multiplier for incompressibility. 
On a closed surface with $\partial M = \varnothing$, the stationary conditions with respect to velocity $\bu$ and pressure $p$ yield the incompressible, steady Stokes equations on a 2D Riemannian manifold:
\begin{align}\label{eqn: stokes flow}
\begin{split}
    2 \mu ~ \bar \partial^{\cvr *} \delbar \bu + \mathrm{grad}~ p - \alpha ~\mathrm{div}^{\cvr} \bQ &=  0,\\
    \text{div } \bu &= 0.
\end{split}
\end{align}
The curvature contributes to the viscous term as $-2 \bar \partial^{\cvr *} \delbar \bu = (\blp + K) \bu$.
This result can be deduced from the known approach based on Riemannian geometry that results in the relation $-2 \mathcal{K}^* \mathcal{K} \bu =( \blp + K + \grad~\div)\bu $ \cite{gilbert_geometric_2023, chan_formulation_2017, samavaki_navierstokes_2020}. 
Additionally, in \ref{sec: complex derivation}, we provide an alternative derivation within the framework of complex manifolds using complex differential forms.

It is worth noting that the presence of a Killing vector field indicates symmetries in the geometry, such as the rigid body rotation on a rotationally symmetric surface.
In such case, the Stokes equations might lack full rank, leading to non-unique solutions with its kernel represented by the Killing vector field. 
In our study, we select the least-norm solution as the canonical one to the Stokes equations.
This choice is equivalent to projecting out all modes of Killing fields in the \(L_2\) sense.
The treatment is justified by the following.  
In theory, the $L_2$ Killing component of the velocity does not contribute to energy dissipation, and as a result, it remains conserved in the presence of any non-zero inertia. 
This is an instance of Noether's theorem, which states that the presence of continuous spatial symmetry, represented by a Killing vector field, implies the conservation of momentum.
Therefore, in such ideal case, the correct solution is the least-norm solution with the addition of the Killing mode in the initial condition.
However, in practical situations, particularly when considering an interfacial fluid surface that separates two bulk viscous fluids, the Killing mode will eventually dissipate energy through frictional interactions with the substrate\footnote{Interactions with the bulk fluid are not considered in this study.}. 
In this case, the system should converges to the least-norm solution over time.
In Sec.~\ref{sec:semi-lagrangian}, we will elaborate on the specific implementation of the Killing projection.

\subsubsection{Nematodynamics}\label{sec: continuous nematodynamics}
Modeling nematodynamics involves studying the dynamics of nematic liquid crystals, which possess both fluid-like and orientational order properties. 
In this study, we are particularly interested in the sharply aligned limit of the active nematic field.
This sharply aligned phase arises when the microscopic rotational diffusion of the nematic molecules is negligible \cite{gao_analytical_2017}.
Under these conditions, the nematodynamics of the macroscopic order parameter becomes identical to the dynamics of individual microscopic nematic filaments.
A commonly used approach is to utilize an advection-diffusion equation, where advection accounts for the flow-induced alignment and rotation of the nematic director field, while diffusion represents the smoothing of the director due to random molecular interactions \cite{berisedwards1994}. 

The relaxation or diffusion part of the nematodynamics equation involves the nematic Bochner Laplacian established in Sec.~\ref{sec: laplacian and diffusion}.
The constraint $|[\bq]| = 1$ imposes the sharply-aligned phase.
However, according to the Poincaré-Hopf theorem, zeros of the director field are guaranteed unless the system is on a torus with a vanishing Euler characteristic $\chi = 0$ (\cf Sec.~\ref{sec: continuous topology}). 
As a result, the hard constraint $|[\bq] |=1$ is in general not permissible for a continuous vector field. 
To address this constraint, we adopt a Ginzburg-Landau formulation and introduce a weak constraint through a penalizing energy term $\mathcal{G}([{\bq}]) = \mathcal{G}(|[\bq]|) = (|[\bq]|^2 - 1)^2 / (4\varepsilon^2)$, where $\varepsilon$ is referred to as the coherence length scale around the defects \cite{Pismen1999VorticesIN}. 
This combination of the Ginzburg-Landau potential and the one-constant Frank-Oseen energy, $\mathcal{G}([{\bq}]) + |\nabla^L [\bq] |^2$, is commonly referred to as the Leslie-Ericksen energy, whose variation induces a Ginzburg-Landau modulated diffusion \cite{ericksen1989liquid, de1993physics, virga2018variational, leslie1968some}.
Note that the nematodynamics in the close vicinity of defects is not captured by the present model and requires more detailed analysis \cite{canevari2018defects}.

The advection of nematic molecules is captured by a combination of the Lie transportation and the ``rotation-only'' Lie transportation.  Lie transportation, as described in Sec.~\ref{sec: lie derivative}, models how the nematic director field is passively transformed by the flow generated by a velocity field.  The rotation-only Lie transportation factors out the stretching part of the Lie transportation, and thus a nematic director is only rotated while keeping its magnitude.

By decomposing the velocity gradient into symmetric ($\bE = \mathcal{K}\bu$, representing strain rate) and skew-symmetric ($\bW = \cvr \bu - \bE$, representing vorticity) components, we can express the nematic Lie derivative (\cf Eqn.~\eqref{eqn: Lie derivative as derivative of pullback} and Sec.~\ref{sec: Lie derivative as covariant derivative}) as $\Lie_{\bu} [\bq] = 2 \bq \Lie_{\bu} \bq$, where $\Lie_{\bu} \bq = \nabla_{\bu} \bq  - \bE \bq - \bW \bq$.
The vorticity $\bW$ and the parallel transport $\cvr_{\bu}$ contribute to infinitesimal rotational transformations that preserve length, whereas the strain rate $\bE$ affects the molecules by aligning them with the flow and causing axial stretching or compression.

A rotational Lie derivative, $\Lie^{\operatorname{rot}}_{\bu}$, also known as the Jaumann or corotational derivative, specifically focuses on the rotation effect induced by the flow, (\eg  $\Lie^{\rot}_{\bu} \bq = \nabla_{\bu} \bq - \bW \bq$ for vector fields \(\bq\)).
Like the classical Lie derivatives (Sec.~\ref{sec: lie derivative}), the rotational Lie derivative \(\Lie^{\rot}_{\bu}\) is defined through a pullback operation.  The only difference is that this pullback is only the rotation component of the classical pullback operator.
Recall Eqn.~\eqref{eqn: continuous pullback} that if \(\varphi\) is a flow map with deformation gradient \(\bF\), the pullback of a vector field and respectively a nematic field is given by \(\varphi^*\bq = \bF^{-1}\bq\), \(\varphi^*[\bq] = [\bF^{-1}\bq]\) and \(\varphi^*\bQ = \bF^{-1}\bQ\bF^{-\top}\).
Now, let \(\bR\) denote the rotational component of \(\bF\) in the polar decomposition \(\bF = \bR\bU\).
Define the \emph{rotational pullback} \(\varphi_{\rot}^*\) on vector fields and nematic fields by
$\varphi_{\rot}^* {\bq} \coloneqq \bR^{-1} \bq$, $\varphi_{\rot}^* {[\bq]} \coloneqq [\bR^{-1} \bq]$, and $\varphi_{\rot}^* \bQ \coloneqq \bR^{-1} \bQ \bR^{-\top}$.
Via this rotational pullback operator, define the Jaumann derivative as \(\Lie_{\bu}^{\rot}\coloneqq {\frac{\partial}{\partial t}}\circ\varphi_{\rot}^*\), where \(\varphi\) is the infinitesimal flow map generated by \(\bu\).
In terms of the gradients \(\nabla\bu\), \(\bE\), \(\bW\) of \(\bu\),
\begin{equation}
    \Lie_{\bu}^{\rot}\bq = \nabla_{\bu}\bq - \bW\bq,\quad
    \Lie_{\bu}^{\rot}[\bq] = 2\bq\Lie_{\bu}^{\rot}\bq,\quad
    \Lie_{\bu}^{\rot}\bQ = \nabla_{\bu}\bQ - \bW\bQ + \bQ\bW.
\end{equation}

The combination of the Jaumann derivative with the traditional Lie derivative in a weighted manner, denoted as $\Lie^{\lambda}_{\bu} \coloneqq \lambda \Lie_{\bu} +  (1-\lambda) \Lie_{\bu}^{\rot}$, $\lambda \in [0, 1]$, allows for the modeling of spheroidal objects with finite ellipticity.
The parameter $\lambda \in [0, 1]$ allows for a linear transition from considering the full velocity gradient at $\lambda = 1$ to only the spin tensor at $\lambda = 0$. 
In the literature, this non-dimensional parameter $\lambda$ is referred to as the tumbling parameter or Bretherton's constant, which determines the dominance of alignment in extensional flow versus tumbling in rotational flow  \cite{bretherton1962, thijssen_active_2020}.

In addition, nematic molecules are inextensible when subjected to flow. 
During advection, the magnitude of the order parameter is transported like a scalar field without undergoing stretching.
Therefore, we project out the stretching component in the Lie derivatives \(\Lie_{\bu}^\lambda\).  Explicitly the projected Lie derivatives are given by
\begin{align}\label{eqn: jeffery equation}
\begin{split}
    &\mathcal{P} (\Lie^{\lambda}_{\bu} \bq) \coloneqq \Lie^{\lambda}_{\bu} \bq - \frac{1}{|\bq|^2} \langle \Lie^{\lambda}_{\bu} \bq, \bq \rangle \bq =  \nabla_{\bu} \bq - \bW {\bq}  - \lambda \bE {\bq} +  \frac{\lambda}{|\bq|^2}\langle \bE {\bq}, \bq \rangle \bq, \\
    & \mathcal{P}^L (\Lie^{\lambda}_{\bu} [\bq])  \coloneqq \Lie^{\lambda}_{\bu} [\bq] - \frac{1}{|[\bq]|^2} \langle \Lie^{\lambda}_{\bu} [\bq], [\bq] \rangle [\bq] =    2 \bq~  \mathcal{P} (\Lie^{\lambda}_{\bu} \bq),\\
&\mathcal{P}^{E} (\Lie^{\lambda}_{\bu} \bQ) \coloneqq \Lie^{\lambda}_{\bu} \bQ - \frac{1}{|\bQ|^2} \langle \Lie^{\lambda}_{\bu} \bQ, \bQ \rangle \bQ = \cvr_{\bu} \bQ -  (\lambda \bE + \bW) \bQ - \bQ  (\lambda \bE - \bW) + \frac{2\lambda}{\vert \bQ \vert} \langle \bE , \bQ \rangle \bQ.
\end{split}
\end{align}
Here we normalize $\bQ$ based on the Frobenius matrix product $\langle \cdot, \cdot \rangle$. 
It is equally valid to use the nematic norm $\langle \cdot, \cdot \rangle_{\mathcal{Q}}$ introduced in Eqn.~\eqref{eqn: nematic Q dirichlet energy}.
Projections based on both norms are equivalent because $\det(\bQ) = 0$, meaning $|\bQ| = |\bQ|_{\mathcal{Q}}$, and $2 \det(\bQ, \bQ') = \det(\bQ, \bQ)' = \det(\bQ)' = 0$, indicating $\langle \bQ', \bQ \rangle = \langle \bQ', \bQ \rangle_{\mathcal{Q}}$.
In the literature, the generalized isometric advection equation based on Eqn.~\eqref{eqn: jeffery equation} is known as the Jeffery equation \cite{jeffery1922ellips}.

By combining Jeffery advection with the Ginzburg-Landau diffusion, we obtain the nematodynamics equations governing the evolution of the surface nematic:
\begin{align}
\begin{split}
    &\dot {[\bq]} + \mathcal{P}^L(\Lie_{\bu}^\lambda[\bq])  = \frac{1}{\eta} \left (\blp^L [\bq] - \frac{\delta \mathcal{G}}{\delta [\bq]} \right), \\
&\dot {\bQ} + \mathcal{P}^E(\Lie_{\bu}^\lambda\bQ) =\frac{1}{\eta} \left ( \blp^E \bQ - d \mathcal{V} \left \llbracket \frac{\delta \mathcal{G}}{\delta [\bq]} \right \rrbracket \right),
\end{split}
\end{align}
where 
\begin{align}
    \frac{\delta \mathcal{G}([\bq])}{\delta [\bq]} = \frac{1}{\varepsilon^2}(| [\bq] |^2 - 1) [\bq] ,
\end{align}
and $\eta$ is the co-called rotational viscosity. 
Although not explicitly apparent, the Gaussian curvature $K$ plays an implicit role in the Bochner Laplacian through the Levi-Civita connection \cite{gilbert_geometric_2023, knoppel_globally_2013}.

\subsubsection{System of equations}\label{sec: system of equations}
We summarize and non-dimensionalize the governing equations for the hydrodynamics of an active nematic fluid film. Using the domain length scale, denoted as $r$, as the characteristic length scale and the time scale $\tau = \mu |\alpha|^{-1}$, we can scale all variables and differential operators. This leads to the dimensionless system given by:

\begin{subequations}\label{eqn: dimensionless system}
\begin{equation}\label{eqn: dimensionless system A}
( \blp +  K)  \bu -  {\text{grad}}~  p  +  \mathrm{div}^{\cvr} \bQ = 0, \quad  \text{div}~ \bu = 0,
\end{equation}
\begin{equation}\label{eqn: dimensionless system B}
\dot{[\bq]} + \mathcal{P}^L \Lie^{\lambda}_{ \bu}  [\bq] = \frac{1}{ \Pe } \left (\blp^L [\bq] -  \frac{1}{\epsilon^2}(| [\bq] |^2 - 1) [\bq] \right ),
\end{equation}
\end{subequations}
where $\lambda \in [0, 1]$ is the tumbling parameter; $\Pe = |\alpha| \eta r^2 \mu^{-1}$ is the active Peclet number, which measures the influence of activity-driven Jeffery advection compared to nematic relaxation; $\epsilon = \varepsilon / r$ characterizes the size of a defect core by comparing the coherence length to the system size.
Throughout the rest of the study, we solve the non-dimensional version of the system. 
In the numerical results presented below, we assume the limit $\epsilon \rightarrow 0$ and adopt a normalization procedure to account for the Ginzburg-Landau term.

\section{Discretization and algorithm \label{sec:methods}}
In this section, we will explain the construction of a discrete system for modeling an active fluid on a 2D triangular manifold mesh. 
The mesh serves as a connected and embedded representation of a continuous 2D manifold, comprising vertices, edges, and faces. 

When discretizing Eqn.~\eqref{eqn: dimensionless system} on a manifold, we need to discretize surface differential operators.
We begin by constructing discrete differential operators for scalar functions using linear finite elements.
Every vertex and face is assigned with a tangent space, over which vector fields and nematic fields are defined.  
Derivatives of vector and nematic fields require the Levi-Civita connection.
By leveraging the Levi-Civita connection, we obtain both the vector and nematic Bochner Laplacian by constructing a discrete Dirichlet energy and taking its variation. 
By a discrete version of the Gauss--Bonnet Theorem, we also obtain Gaussian curvatures on vertices and faces.

Using these discrete differential operators, we solve Eqn.~\eqref{eqn: dimensionless system} by time splitting (\cf Alg.~\ref{alg:outline}). 
At each time step, the active force is calculated by taking the covariant divergence of the stress tensor induced by the nematic field.
This active force gives rise to a fluid velocity field, which is obtained by solving the discrete Stokes equations (Eqn.~\eqref{eqn: dimensionless system A}) through Augmented-Lagrangian (AL) iteration.
Next, the nematodynamics equation (Eqn.~\eqref{eqn: dimensionless system B}) evolves the nematic field with the advection by the Stokes flow velocity field, as well as the elastic diffusion. 
The advection and diffusion terms are separated in a splitting procedure.
We first compute the evolution of the nematic field Lie-advected by the velocity using an explicit semi-Lagrangian (sL) scheme. 
The one-step diffusion, based on the Bochner Laplacian operator, is solved using the implicit Euler method to ensure unconditional stability. 
The projection that removes the stretching component of the Lie derivative and the constraint imposed by the Ginzburg-Landau term are consolidated into a single normalization step.
For completeness, we also include a discrete counterpart of the theory of Sec.~\ref{sec: continuous topology} about curvature and defects.
The resulting algorithm enables us to determine the locations and the charges of the topological defects of the nematic field.

\begin{algorithm}
\caption{Hydrodynamics of active nematics}
\label{alg:outline}
\begin{algorithmic}[1]
\Inputs{Surface $M$, curvature $K$, time step $\Delta t$}
\Initialize{Nematic field $\bQ = \mathcal{V}([\bq])$}
\For{$t = 0$ to $T$}
\State \Call{Stokes flow}{$\bQ$}
  \State  \hspace*{2em}{$ \bff \gets \mathrm{div}^{\cvr} \bQ $} 
    \State  \hspace*{2em}{ $
    \bu, p \gets  \Call{Solve}{
    (\blp + K) \bu = \text{grad } p - \bff  ,\text{div } \bu = 0 
    } $} \Comment{ AL iteration (Sec.~\ref{sec: augmented lagrangian})}
\State \Call{Nematodynamics}{$\bu$, $\bQ$}                
        \State \hspace*{2em} {$\bQ \gets  \Call{Solve}{\dot \bQ + \mathcal{P}^E \Lie^{\lambda}_{\bu} \bQ = 0}$} \Comment{Explicit sL (Sec.~\ref{sec:semi-lagrangian})}
        \State \hspace*{2em} {$[\bq] \gets \mathcal{V}^{-1}(\bQ) $}
        \State \hspace*{2em} {$[\bq] \gets  \Call{Solve}{\dot {[\bq]} = \Pe^{-1} \blp^L [\bq]}$} \Comment{Implicit Euler}
        \State \hspace*{2em} {$\bQ \gets \mathcal{V}(\Call{Normalize}{[\bq]})$}
\State \hspace*{2em} $t \gets t + \Delta t$
\EndFor
\end{algorithmic}
\end{algorithm}

In summary, the method presented in this work offers several distinctive features that contribute to the effectiveness and versatility of the method in modeling and analyzing the behavior of active nematics on curved surfaces:
\begin{itemize}
\item The calculations are carried out on a 2D triangular mesh, rather than a volumetric grid, allowing for efficient and accurate representation of the system.
\item Tensor fields and differential operators are intrinsically represented using the discrete complex line bundle.
This representation, in contrast to the common approach of applying extra tangentiality constraints on Euclidean tensors and operators, allows better accuracy and lower computational cost.
\item A generalized semi-Lagrangian method is employed to handle the Lie derivative for both vector and tensor quantities, thereby providing a general procedure for the advection of $k$-atic directors.
\item We utilize isomorphic representations of the nematic field through the Veronese map.  
This allows us to adapt to the more convenient representation at each stage of the algorithm.
\end{itemize}

\subsection{Discrete manifold and choice of variable space \label{sec: mesh setup}}
In this section, our goal is to establish notations for representing a 2D manifold using a discrete mesh and navigating its local topology. 
Our focus here lies in developing a common language for tensor calculus on a discrete manifold, which forms the basis for discretizing the system of equations. 

\subsubsection{Representation and navigation of discrete 2D manifolds}\label{sec: mesh representation}
\begin{figure}[htbp]
    \centering
    \includegraphics[width = 4.5 in]{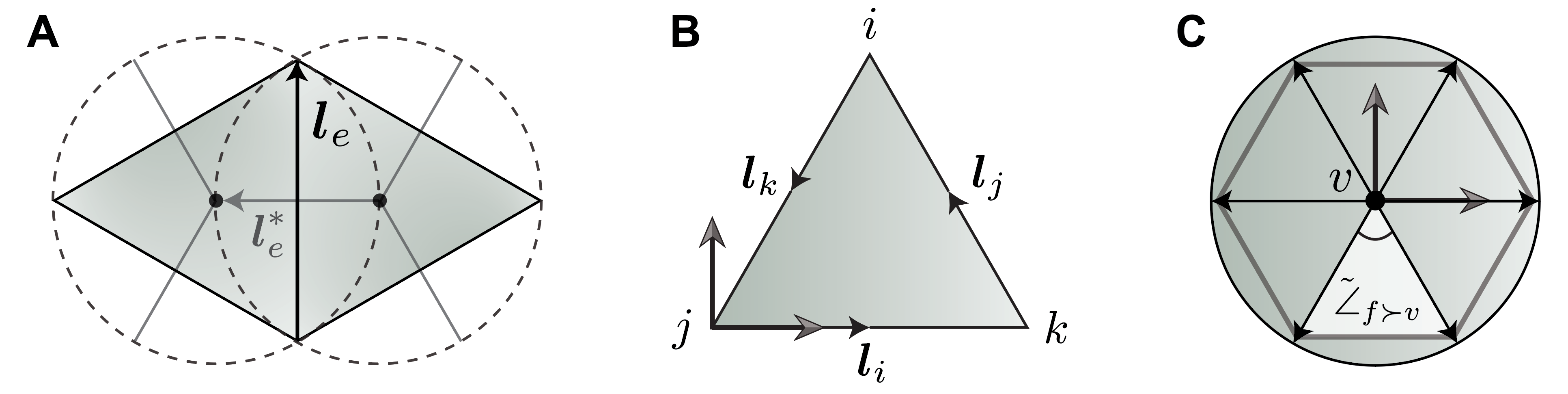}
    \caption{Schematics illustrating key components of a 2D triangular mesh: A) Voronoi dual mesh is constructed by circumcenters of the triangular mesh, defining the weight of each primal element. B) Tangent space associated with the faces of the mesh, denoted as $T_fM$. C) Vertex tangent space $T_vM$ via the geodesic polar map.}
    \label{fig:schematics}
\end{figure}

A closed, oriented triangular mesh $M$ of arbitrary genus $g$ consists of the sets of vertices, edges, and faces, denoted as $\{\mathfrak{V, E, F}\}$.
Let $|\mathfrak V|$, $|\mathfrak E|$, and $|\mathfrak F|$ denote the numbers of vertices, edges, and faces, respectively.
To indicate the local topology of the mesh, we use the symbols $\prec$ and $\succ$, which signify the relation ``incident to". The orientation of the symbol indicates the dimensionality relationship. 
For example, for incident $v \in \mathfrak V$, $e \in \mathfrak E$, and $f \in \mathfrak F$, we write $v \prec e \prec f$.
When using the summation sign, the iterating variable is placed first. 
For example, $\sum_{f \succ v}$ means summing over faces $f$ that are adjacent to a given vertex $v$.
The directed edge vector associated with the edge $e \in \mathfrak E$ in face $f \in \mathfrak F$, where $e \prec f$, is denoted as $\bl_{e \prec f}$. 
This vector is positively-oriented (\ie counterclockwise) in face $f$.
In particular, this edge has opposite orientations in the two faces that share it.
Similarly we can also denote the edge vector associated with a vertex as $\bl_{v \prec f}$, which has the positive orientation of pointing outward from the vertex.
Lastly, the angle of the interior corner of a triangle \(f\) at a vertex \(v\prec f\) is denoted by \(\angle_{v\prec f}\), \(\angle_{f\succ v}\), \(\angle_{e\prec f}\), or \(\angle_{f\succ e}\), where \(e\) is the edge opposite to vertex \(v\) across triangle \(f\).

Using the circumcenters of the mesh faces, we can construct a \emph{Voronoi dual mesh}. 
In this dual mesh, each edge, known as the dual edge (illustrated by Fig.~\ref{fig:schematics}A), orthogonally bisects the corresponding primal edge in the primal mesh.
Note that the dual edge is intrinsically straight, but it may be kinked when the mesh is embedded due to the dihedral angle formed by neighboring faces.
We denote the directed dual edge as $\bl^*_{e \prec f} = w_e \Img \bl_{e \prec f}$, where $w_e = {|\bl_e|}/{|\bl^*_e|}$ is the length ratio between the dual edge and the primal edge referred to as the \emph{cotangent weight} \footnote{since \(w_e = {\frac{1}{2} }(\cot(\angle_{e\prec f_1})+\cot(\angle_{e\prec f_2}))\), where \(f_1\neq f_2\succ e\) are the two incident faces of \(e\).}, and $\Img$ representes $90^\circ$ counter-clockwise rotation in face $f$.
Analogous to the continuous notation that $C^{\infty}(M)$ denotes the space of continuous functions on $M$, we use $C_{\mathfrak V}(M) = \{h\colon\mathfrak{V}\to\Real\}$ to denote the space of functions defined over vertices.
The function evaluation at vertex $v \in \mathfrak V$ is denoted as $h_v$.

The notion of a tangent space to a mesh includes the tangent space $T_vM$ at a vertex $v$ and the face tangent space $T_fM$ at a face $f$.
A vector field defined on faces is a \emph{face section} of the tangent bundle, $\Gamma_{\mathfrak F}(TM)$, and similarly, a vector field defined on vertices is a \emph{vertex section} denoted as $\Gamma_{\mathfrak V}(TM)$.
A vector basis section is represented by $\Base \in \Gamma(TM)$, where the basis vector at a vertex $v$ is denoted as $\Base_v$, and at a face $f$ as $\Base_f$.
Given a basis section, a vector field can be represented by a complex-valued function. 
For example, a face-based velocity field, $\bu \in \Gamma_{\mathfrak F}(TM)$, can be represented as $\coeff u : \mathfrak F \rightarrow \mathbb{C}$.
The same representation holds for a nematic field with its basis section denoted as $[\Base] \in \Gamma(L)$.
Specifically, on the face tangent space $T_fM$, we start by selecting an arbitrary positively-oriented normalized edge vector (\eg the first edge $e_1$ in the data structure) as the basis, denoted as $\Base_f = \bl_{e_1 \prec f} / |\bl_{e_1 \prec f}| \in T_fM$ and $[\Base]_f = [\Base_f] \in L_f$ (\cf Fig.~\ref{fig:schematics}B).
Then, we can associate all vectors and nematic director in this tangent space with a complex number.
In particular, we represent the phase angle of the oriented edge $\bl_{e \prec f} = \coeff l_{e \prec f} \Base_f $ as $\arg(\bl_{e \prec f}) \coloneqq \arg (\coeff l_{e \prec f}) $. 
This phase angle is used to compute the discrete Levi-Civita connection in Sec.~\ref{sec: discrete Levi-Civita}.

On a vertex tangent space $T_vM$, which is cone-shaped, we can construct an intrinsic representation referred to as the \emph{geodesic polar map} (\cf Fig.~\ref{fig:schematics}C).
To achieve this, we first need to flatten the cone.
However, due to inherent curvature, a cone generally cannot be flattened perfectly due to angle defects or excesses.
In other words, the sum of all corner angles of the faces incident to the vertex does not equal $2\pi$ (\ie $\sum_{f \succ v} \angle_{f \succ v} \neq 2\pi$).
To perform an intrinsic projection, we scale the corner angles by the ratio $r_v = (2\pi) / \sum_{f \succ v} \angle_{f \succ v}$. 
By applying this scaling factor to all interior angles at vertex \(v\), we obtain a \emph{geodesic angle} for each corner, denoted as $\tilde{\angle}_{f \succ v} = r_v \angle_{f \succ v}$.
Similar to the face tangent space, we choose an arbitrary positively-oriented edge vector $\bl_{e_1\prec v}$ at the vertex as the vector basis and its corresponding nematic basis, represented by $\Base_v \coloneqq \bl_{e_1 \prec v} / \vert \bl_{e_1 \prec v}\vert \in T_v M $ and $[\Base]_v = [\Base_v] \in L_v$.
We accumulate the geodesic corner angles and obtain the phase angle $ \arg(\bl_{e \succ v}) = \arg(\coeff l_{e \succ v} )$ for each edge $e$ in this vertex polar plane. 
Consequently, the edge vector in $T_vM$ can be represented by a complex number $\bl_{e \succ v} = \coeff l_{e \succ v} \Base_v$.

\subsubsection{Levi-Civita connection}\label{sec: discrete Levi-Civita}

We equip the discrete manifold with a Levi-Civita connection (\cf Sec.~\ref{sec: continuous connection}) by locally constructing the discrete parallel transport. 
This includes the parallel transport from the tangent space of one vertex to its neighboring vertex, from a face to its neighboring face, as well as from a vertex to its neighboring face.

\begin{figure}[htbp]
    \centering
    \includegraphics[width=3.5 in]{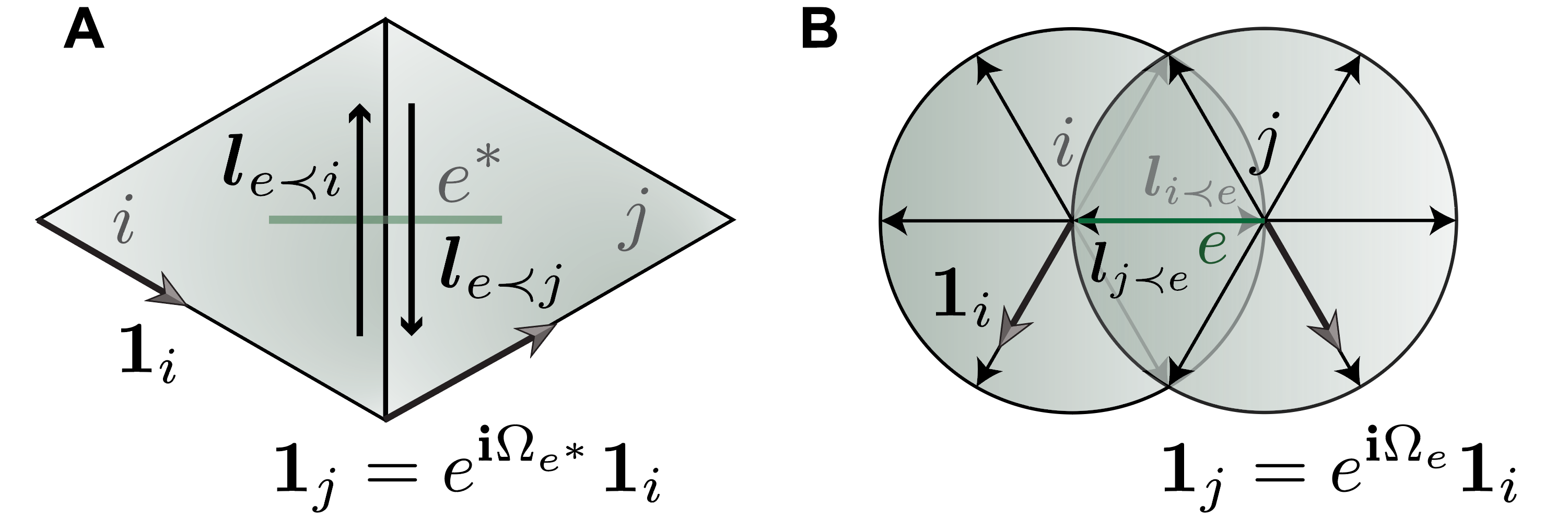}
    \caption{Schematics depicting intrinsic parallel transport on a 2D triangular mesh: A) face-face transportation by flattening the dihedral angle within a two-triangle stencil. B) vertex-vertex transportation preserves angles along geodesic paths within the geodesic polar map.}
    \label{fig:levi civita}
\end{figure}

The face-face and vertex-vertex transportations can be defined intrinsically (\ie invariant under isometric deformation).
Therefore, parallel transport of a vector $\bu$ from face $f_i$ to face $f_j$ along dual edge $e^*$, $e \prec f_i, f_j$, can be conceptualized as if the dihedral angle at $e$ were flattened (\cf Fig.~\ref{fig:levi civita}A).
In this context, parallel transport simplifies to a basis change across face $f_i$ to face $f_j$,
\begin{align}\label{eqn: face-face levi-civita}
\Pi_{i \rightsquigarrow j}: T_{f_i}M \rightarrow T_{f_j}M, \quad \Pi_{i \rightsquigarrow j} (\bu_i) = \Pi_{i \rightsquigarrow j} ( \coeff{u}_i \Base_i) =  e^{- \Img \Omega_{e^*}} \coeff{u}_i \Base_j,
\end{align}
where the angle-valued connection along the dual edge, $ \Omega_{e^*} = \arg(\bl_{e \prec f_j}) - \arg(\bl_{e \prec f_i}) - \pi $, accounts for the basis change, $\Base_j = e^{\Img \Omega_{e^*}} \Base_i$ (\cf  Fig.~\ref{fig:levi civita}A and the continuous theory discussed around Eqn.~\eqref{eqn: covariant derivative}).

Analogous to the continuous theory that parallel transport can be achieved by preserving the angle along a geodesic path (\cf Sec.~\ref{sec: continuous connection}), the vertex-vertex transportation is intrinsically defined by maintaining the angle with respect to the connecting edge $e$ between neighboring vertices $v_i, v_j \prec e$, illustrated by Fig.~\ref{fig:levi civita}B.
As seen in the continuous theory (\cf Sec.~\ref{sec:Levi-Civita connection on the nematic bundle}), the nematic field requires amplifying the Levi-Civita connection on tangent bundle $\Omega$ to $2 \Omega$.
Subsequently, we can express the parallel transport of a nematic director $[\bq]_i \in L_{v_i}$ to $L_{v_j}$ along $\bl_e$ as 
\begin{align}\label{eqn: vertex-vertex levi-civita}
  \Pi_{i \rightsquigarrow j}: L_{v_i} \rightarrow L_{v_j}, \quad 
  \Pi_{i \rightsquigarrow j} ([\bq]_i) = \Pi_{i \rightsquigarrow j} ( \coeff{[q]}_i [\Base]_i) =  e^{- \Img 2 \Omega_{e}} \coeff{[q]}_i [\Base]_j.
\end{align}
Here, the primal edge connection, $ \Omega_{e} = \arg(\bl_{e \succ v_j}) - \arg(\bl_{e \succ v_i}) - \pi $, quantifies the basis change across $T_{v_i} M$ and $T_{v_j} M $ (\cf Fig.~\ref{fig:levi civita}B).

The vertex-face transportation is achieved using the $\Real^3$ embedding. 
Given the embedding of the mesh and the realization of the tangent vector in $\mathbb{R}^3$, we can utilize a 3D dihedral rotation to map from one vertex tangent space to a neighboring face tangent space and vice versa (\cf the continuous theory in Sec.~\ref{sec: continuous connection}). 
To illustrate, consider a vertex unit normal, $\bn_v \in T_v^\perp M$,$ |\bn_v| = 1$, and a face unit normal, $\bn_f \in T_f^\perp M$, $|\bn_v| = 1$. 
The dihedral rotation from $\bn_v$ to $\bn_f$ is defined as the rotation \(\bR_{vf}\) of with angle $\cos^{-1} \langle \bn_v, \bn_f \rangle$ about the axis $\bl = {\bn_v \times \bn_f}/{|\bn_v \times \bn_f|}$.
Define \(\Pi_{v\rightsquigarrow f}\coloneqq\bR_{vf}: T_{v}M \rightarrow T_{f}M\).
Note that the definition of $\bn_v$ is not unique. 
In practice, we use an angle-weighted vertex normal $\bn_v \coloneqq (\sum_{f \succ v} \tilde{\angle}_{f \succ v} \bn_f ) / (2 \pi)$.

\subsubsection{Space of system variables and operator mappings\label{sec: Space of system variables and operator mappings}}
In this section, we specify the choice of space where the discretization variable resides, and the mapping spaces imposed by the differential operators.
In Secs.~\ref{sec:semi-lagrangian}-\ref{sec: discrete laplacian}, we will recapitulate the specific definitions of the operators.

As we mentioned in the previous section, it is convenient to define tangent spaces over both the vertices and faces of the mesh.
In our approach, we discretize the Stokes equations on the faces of the mesh and the nematodynamics equation on the vertices. 
Specifically, we work with a face-based fluid velocity denoted as $\bu \in \Gamma_{\mathfrak{F}}(TM)$ and a vertex-based nematic director represented as $[\bq] \in \Gamma_{\mathfrak{V}}(L)$, alongside the corresponding matrix representation $\bQ = \bq \otimes \bq \in \Gamma_{\mathfrak{V}}(E)$.

Subsequently, we develop the discrete differential operators associated with these choices. 
In the context of the Stokes equations, we construct the Bochner Laplacian, denoted as $\blp : \Gamma(TM) \rightarrow \Gamma(TM)$, utilizing the face-face Levi-Civita connection. 
The gradient operator, $\grad: C_{\mathfrak{V}}(M) \rightarrow \Gamma_{\mathfrak{F}}(TM) $, maps scalar vertex data to vector data on faces. 
The divergence operator, the adjoint of the gradient, maps vectors from faces to scalar measurements at vertices, $\div:  \Gamma_{\mathfrak{F}}(TM) \rightarrow C_{\mathfrak{V}}(M)$.
Similarly, in the context of the nematodynamics equation, we establish the nematic Bochner Laplacian as $\blp^L [\bq] : \Gamma_{\mathfrak{V}}(L) \rightarrow \Gamma_{\mathfrak{V}}(L)$, using the vertex-vertex Levi-Civita connection.
We also introduce a vertex-based Lie derivative, denoted as $\Lie_{\bu}: \Gamma_{\mathfrak V} (E) \rightarrow \Gamma_{\mathfrak V}(E)$, to advect the nematic director.

\subsection{Semi-Lagrangian Lie advection and nematodynamics}\label{sec:semi-lagrangian}

The sL method is an explicit space-time integrator for advection equations similar to an upwinding scheme \cite{hundsdorfer_numerical_2003}.  
Here we introduce the sL method generalized to any Lie advection equation for tensor fields as follows. 
Given a velocity field, we construct a backward flowmap by tracing the velocity field upwind. 
This flowmap is used to pullback the tensor field from the previous time step as the update rule.
The principle can be applied to tensors of arbitrary $(r, s)$ type (\eg the valence-$k$ tensor that represents the $k$-atic field \cite{giomi2022hydrodynamic}), serving as a discrete counterpart to the continuous theory discussed in Sec.~\ref{sec: lie derivative}.
The resultant scheme, summerized in Alg.~\ref{alg:semi-lagrangian}, captures the geometric structure of nematic advection and enhance numerical stability.

Concretely, given a time-independent velocity field $\bu\in \Gamma_{\mathfrak V}(TM)$, we construct a finite-time backward flowmap $\Psi_{\Delta t}$ by integrating the ODE, $ \partial_\tau \Psi_{\tau} = -\bu(\Psi_\tau)$, $\tau \in [0,\Delta t]$, with the initial condition $\Psi_0 = \mathrm{id}_M$.
In practice, we represent $\Psi_{\tau}$ and $\bu$ as \(\Real^3\)-valued functions using surface embeddings and implement the integration using RK4. 
To evaluate \(\bu\circ\Psi_{\tau}\), we use a closest-point projection followed by a linear interpolation in a triangle.
Recall in the continuous setting that Lie advection $\partial_t {[\bq]} + \Lie_{\bu} [\bq] = 0$ is characterized by a constant pullback field $\partial_{t} \varphi^* [\bq] = 0$ (\cf Eqn.~\eqref{eqn: Lie derivative as derivative of pullback}). 
With the backward flowmap \(\Psi_{\Delta t}\) and its gradient $d \Psi_{\Delta t}$, a discrete Lie advection is analogously represented by updating the nematic field $[\bq]$ using its pullback value $ \Psi_{\Delta t}^* [\bq]$. 
The procedure will be made explicit in following subsections.

\subsubsection{Deformation gradient of backward flowmap}

\begin{figure}[htbp]
    \centering
    \includegraphics[width=3.5 in]{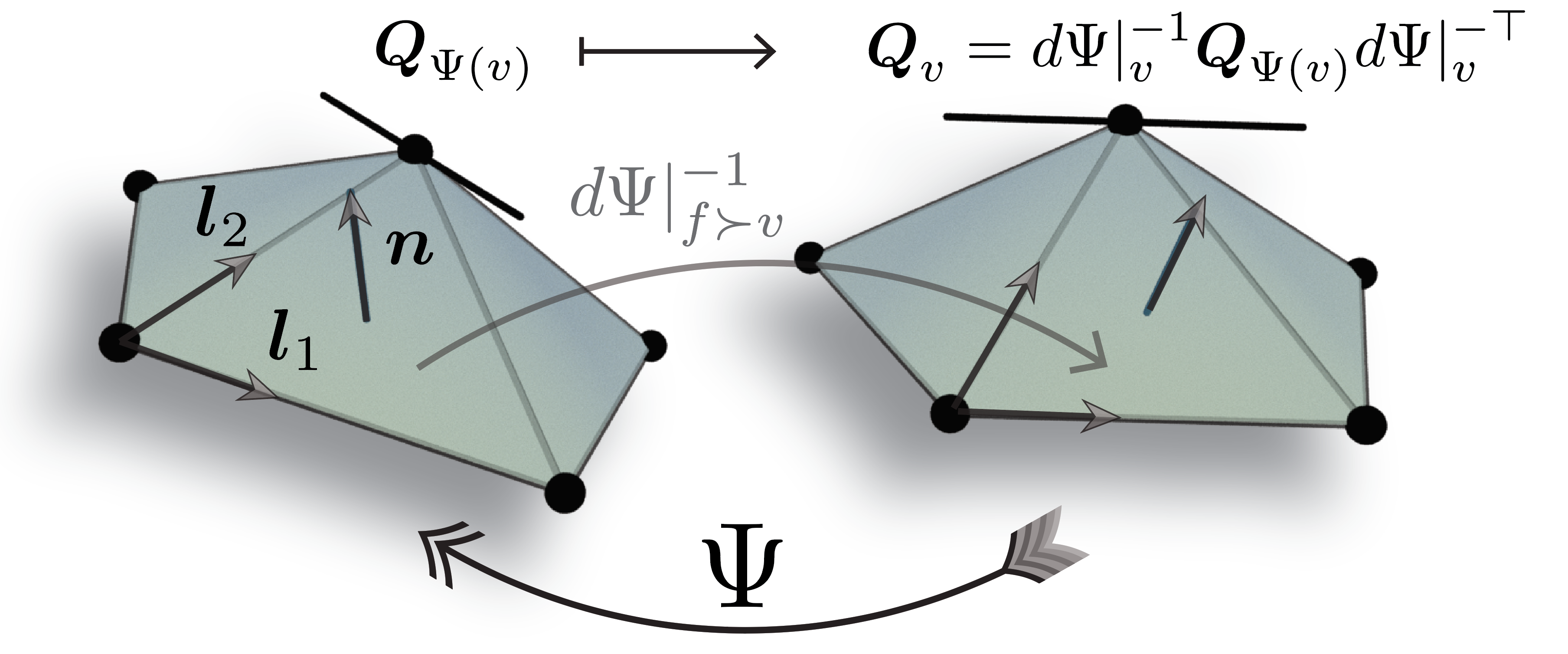}
    \caption{The sL Lie advection is performed on a triangular mesh using a backward finite time flowmap \( \Psi \). 
    This flowmap engenders a piecewise constant deformation gradient \( d \Psi|_{f \succ v} \) defined over the triangle faces.
    By averaging this face-based deformation gradient around the one-ring neighborhood of each vertex, we derive a vertex-based deformation gradient $d \Psi|_{v}$. 
    The nematic Lie advection is modeled by updating $\bQ_v$ with $d \Psi|_v^{-1} \bQ_{\Psi(v)} d \Psi|_v^{-\top}$.}
    \label{fig:lie}
\end{figure}

By evaluating the finite-time flowmap over vertices (\ie $\Psi_{\Delta t}$ such that $M_{\Delta t} = \Psi_{\Delta t} M$), we can construct the deformation gradient $ d \Psi_{\Delta t}|_f$ (realized in $\Real^3$) on each triangle face $f$ as follows:
\begin{align}\label{eqn: discrete deformation gradient}
d \Psi_{\Delta t}: \Gamma_{\mathfrak F}(TM) \rightarrow \Gamma_{\mathfrak F}(T_{\Psi_{\Delta t}} M_{\Delta t}), \quad
d \Psi_{\Delta t} |_f \coloneqq  
    \begin{bmatrix}
         \underset{|}{\overset{|}{\bl_1}} & \underset{|}{\overset{|}{\bl_2}} & \underset{|}{\overset{|}{\bn}}
    \end{bmatrix}_{\Psi_{\Delta t} (f)}
    \begin{bmatrix}
         \underset{|}{\overset{|}{\bl_1}} & \underset{|}{\overset{|}{\bl_2}} & \underset{|}{\overset{|}{\bn}}
    \end{bmatrix}_{f}^{-1}.
\end{align}
These column vectors, illustrated by Fig.~\ref{fig:lie}, represent the embedded edge vectors $\bl \in \Real^3$ and the unit face normal $\bn \in \Real^3$ before and after the backward flowmap $\Psi_{\Delta t}$.
A vertex-based deformation gradient $\bF: \Gamma_{\mathfrak V}(TM) \rightarrow \Gamma_{\mathfrak V}(T_{\Psi_{\Delta t}} M_{\Delta t}) $ is obtained by averaging deformation gradient on incident faces, $\bF |_v = \frac{1}{|f \succ v|} \sum_{f \succ v} d \Psi_{\Delta t}|_{f}$.

\subsubsection{Pullback of nematic field}
Recall that the nematic field can be equivalently expressed using the matrix representation $\bQ = \mathcal{V}([\bq]) \in \Gamma_{\mathfrak V}(TM \otimes TM)$.
With $\bQ$ realized as an $\mathbb{R}^{3 \times 3}$-valued function, its Lie advection, $\partial_t {\bQ} + \Lie_{\bu} \bQ = 0$, with $t \in [0, \Delta t]$, $\bQ|_{t = 0} = \bQ_0$, can be modeled by 
\begin{align}
\bQ_{\Delta t} = \Psi_{\Delta t}^* \bQ_0  = \bF^{-1}  (\bQ_0 \circ \Psi_{\Delta t}) \bF ^{-\top},
\end{align}
where $\bQ_{\Delta t}$ denotes the updated value of $\bQ$ for the subsequent time step.
Because the backward flowmap $\Psi_{\Delta t}$ can land anywhere on $M$, evaluating $\bQ_0 \circ \Psi_{\Delta t}$ requires interpolation, such as trilinear interpolation utilizing the face-barycentric representation of $\Psi_{\Delta t}$.
However, this interpolation may not preserve the rank-1 structure of $\bQ$.
To address this, we perform SVD and and extract the leading component of the interpolated value, ensuring that $\bQ_0 \circ \Psi_{\Delta t} = \bq_{\Psi_{\Delta t}} \otimes \bq_{\Psi_{\Delta t}} $, and thereby $\bQ_{\Delta t}$ remains rank-1.
Lastly, by projecting $\bQ_{\Delta t}$ onto the surface using tangent bases, we recover the intrinsic representation of $\bQ_{\Delta t}$ and its corresponding complex representation $[\bq]_{\Delta t} = \mathcal{V}^{-1}(\bQ_{\Delta t})$.

To model the Jeffery equation $\dot \bQ + \mathcal{P}^{E} (\Lie^{\lambda}_{\bu} \bQ) = 0$ (\cf Eqn.~\eqref{eqn: jeffery equation}), we can include the additional rotational Lie advection $\Lie_{\bu}^{\rot}$ and normalization as follows:
\begin{align}
    \bQ_{\Delta t} = \frac{\lambda \bF^{-1}  (\bQ_0 \circ \Psi_{\Delta t}) \bF ^{-\top} + (1 - \lambda) \bR ^{-1}  (\bQ_0 \circ \Psi_{\Delta t}) \bR ^{-\top}}{|\lambda \bF^{-1}  (\bQ_0 \circ \Psi_{\Delta t}) \bF ^{-\top} + (1 - \lambda) \bR ^{-1}  (\bQ_0 \circ \Psi_{\Delta t}) \bR ^{-\top}|},
\end{align}
where $\bR$ isolates rotation in $\bF$ through polar decomposition $\bF = \bR \bU$. 

As a side remark, in the special case where the transported field $h$ is a scalar function, its advection $\partial_t h +  \Lie_{\bu} h = \partial_t h + \nabla_{\bu} h = 0$ simplies to the traditional sL method, $h_{\Delta t} = \Psi_{\Delta t}^* h_0  = h_0 \circ \Psi_{\Delta t}$.

\begin{algorithm}
\caption{sL method for nematic Lie advection over time interval $[0, \Delta t]$}
\label{alg:semi-lagrangian}
\begin{algorithmic}[1]
\Inputs{Surface mesh $M$, nematic field $\bQ_0 = \mathcal{V}([\bq]_0)$, time step $\Delta t$, flow velocity $\bu$}
\For{each vertex $v \in  \mathfrak V$, %
} 
  \State  \hspace*{2em}{$ \Psi_{\Delta t} \gets  \Call{RK4}{ {d \Psi_\tau}/{d \tau} = - \bu (\Psi_\tau) }$}  \Comment{Backward flowmap}
  \State  \hspace*{2em}{$\bQ_{\Psi_{\Delta t}} \gets \Call{Interp}{\bQ_0 \circ \Psi_{\Delta t}} $}
  \State \hspace*{2em}$\bQ_{\Psi_{\Delta t}} \gets  \Call{SVD}{\bQ_{\Psi_{\Delta t}}} $
\EndFor
\For{each face $f \in  \mathfrak F$} \Comment{Deformation gradient}
  \State  \hspace*{2em}{$ d \Psi_{\Delta t} \gets  [\bl_1, \bl_2, \bn]_{\Psi_{\Delta t}} [\bl_1, \bl_2, \bn]^{-1}$} 

\EndFor
\For{each vertex $v \in  \mathfrak V$, 
}\Comment{Pullback}
  \State  \hspace*{2em}{$ \bF \gets \Call{Avg}{d \Psi_{\Delta t} |_{f \succ v}}  $} 
\State \hspace*{2em} $\bQ_v \gets \bF^{-1} \bQ_{\Psi_{\Delta t}} \bF^{- \top}$
\EndFor
\end{algorithmic}
\end{algorithm}

\subsubsection{Discussion of sL advection scheme}

Although the traditional scalar sL method guarantees unconditional stability, the extension to tensor Lie advection in general does not \cite{nabizadeh_covector_2022}.
The stability depends on the conditioning of the pullback map.
However, by incorporating a diffusion term into the sL advection through time-splitting (\cf Alg.~\ref{alg:outline}) and scaling the system with respect to advection, the time step for advection remains fixed regardless of the P\'eclet number.
Because the diffusion equation, $\dot{[\bq]} = \Pe^{-1} \blp^{L} [\bq]$, is solved using an implicit method (\cf Alg.~\ref{alg:outline}), the algorithm is numerically stable with high P\'eclet number. 
This also highlights some well-known issues associated with time-spliting and the sL method.
First, time-spliting introduces errors, which can be reduced using techniques such as the Strang method \cite{hundsdorfer_numerical_2003}.
Second, sL method based on linear interpolation has significant numerical diffusion.
Improvements can be made by implementing higher-order interpolation and techniques such as the McCormack scheme \cite{hundsdorfer_numerical_2003}.
However, these challenges are common to the sL method and are not unique to the generalization to tensor-valued functions.
Despite these challenges, the current minimal implementation is sufficient for our purpose of studying active nematics on curved surfaces.

\subsection{Gradient and (covaraint) divergence} \label{sec: gradient and covariant divergence}

We used the linear finite element hat function $\Phi_v$ to discretize the gradient for a scalar function $p \in C_{\mathfrak V}(M)$ as follows:
\begin{align}\label{eqn: discrete gradient}
     \grad: C_{\mathfrak V}(M) \rightarrow \Gamma_{\mathfrak F}(TM), \quad A_f (\grad ~p)_f = A_f \sum_{v \prec f} p_v (\grad_f \Phi_v) = \frac{1}{2} \sum_{v \prec f} p_v \Img \bl_{v \prec f},
\end{align}
 where $ \Img \bl_{v \prec f}$ denotes the $90^\circ$ rotation of edge vector $\bl_{v \prec f}$ and $A_f$ is the face area (\cf Sec.~\ref{sec: mesh representation}).

The divergence operator is given by the negated adjoint operation of the gradient.  
The adjoint of Eqn.~\eqref{eqn: discrete gradient} applied to a face-based vector field \(\bu\in\Gamma_{\mathfrak F}(TM)\) is a measure on each vertex \((\grad^*\bu)_v = -{1\over 2}\sum_{f\succ v}\langle\Img\bl_{v\prec f},\bu_f\rangle\), representing the total divergence at vertex \(v\).  Using a reference vertex area measure \(A_v\) we define the divergence operator as
\begin{align}\label{eqn: scalar divergence}
    \div: \Gamma_{\mathfrak F}(TM) \rightarrow C_{\mathfrak V}(M), \quad A_v (\div ~ \bu)_v = - \frac{1}{2} \sum_{f \succ v} \langle \Img \bl_{v \prec f}, \bu_f \rangle. 
\end{align}
Note that the discrete equation involving the divergence operator is often written as in the weak form, where divergence is treated as a measure.  In particular these discrete equations are independent of the specific choice of the vertex area measure \(A_v\).
In cases where pointwise evaluation is needed, such as in visualization, we use $A_v = \sum_{f \succ v} A_f / 3$.

The divergence operator, represented by Eqn.~\eqref{eqn: scalar divergence}, can also be understood as a finite-volume discretization applied to the dual cells of the mesh \cite{hundsdorfer_numerical_2003}.
We can analogously obtain a vertex-based divergence $\div: \Gamma_{\mathfrak V}(TM) \rightarrow C_{\mathfrak F}(M)$ by applying finite volume on the primal mesh.

By incorporating an additional vertex-face parallel transport, $\Pi_{v \rightsquigarrow f}: T_{v}M \rightarrow T_{f}M$, (\cf Sec.~\ref{sec: discrete Levi-Civita}), we can extend such construction to discretize covariant divergence:
\begin{align}\label{eqn: discrete covariant divergence}
    \div^{\cvr}: \Gamma_{\mathfrak V}(TM \otimes TM) \rightarrow \Gamma_{\mathfrak F}(T M), \quad A_f (\div^{\cvr} ~ \bQ)_f = - \frac{1}{2} \sum_{v \prec f} \Pi_{v \rightsquigarrow f} \left [ \bQ_v \cdot  \Pi^{-1}_{v \rightsquigarrow f}( \Img \bl_{v \prec f}) \right] .
\end{align}
The Euclidean dot product, denoted as $\cdot$, denotes a contraction operation on one slot of the $(2, 0)$ tensor (\ie $\bQ \cdot \bq \equiv Q^{ij} q_{j}$).
In the case of $\bQ = \bq \otimes \bq$ being rank-1, the covariant divergence can be explicitly expressed, $A_f (\div^{\cvr} ~ \bQ)_f = - \frac{1}{2} \sum_{v \prec f} \langle \Pi_{v \rightsquigarrow f}(\bq_v) , \Img \bl_{v \prec f} \rangle \Pi_{v \rightsquigarrow f} (\bq_v)$.
In practice, these discrete operators, including those in Sec.~\ref{sec: discrete laplacian}, are organized and computed using sparse matrix multiplication.

\subsection{Bochner Laplacian}\label{sec: discrete laplacian}
Here, we construct the discretization of Bochner Laplacian using the covariant derivative and covariant divergence. 
We will first discuss the covariant derivative and Bochner Laplacian for face-based vector fields and then proceed with a similar construction for vertex-based nematic fields.

Using the face-face parallel transport $\Pi_{i \rightsquigarrow j}, i, j \in \mathfrak F$ across edge $e \prec i, j$ (\cf Eqn.~\eqref{eqn: face-face levi-civita}), discrete directional covariant derivative $\cvr_{\bl^*}$ maps a face-based vector field $\bu = \coeff u \Base \in \Gamma_{\mathfrak F}(TM)$ to an edge-based vector field,
\begin{align}\label{eqn:discrete covariant derivative}
  \cvr_{\bl^*}: \Gamma_{\mathfrak F}(TM) \rightarrow \Gamma_{\mathfrak E}(TM), \quad \vert \bl_e^* \vert (\cvr_{\bl^*} \bu)_e = (\bu_i - \Pi_{i\rightsquigarrow j} (\bu_j) )_{i, j \succ e} =  \left ( (\coeff{u}_j - e^{- \Img \Omega_{e^*}} \coeff{u}_i) \Base_j \right )_{i, j \succ e}.
\end{align}
By taking the $L_2$ norm with the edge area as the diamond area formed by $\bl$ and $\bl^*$, $A_e = |\bl| |\bl^* |$, we can construct the discrete Dirichlet energy as:
\begin{align}
2 \mathcal{F}(\bu) = \llangle \cvr_{\bl^*} \bu,  \cvr_{\bl^*} \bu \rrangle = 
\sum_{e \in \mathfrak E} A_e |\cvr_{\bl^*} \bu|_e^2  
= \sum_{e \in \mathfrak E} \frac{A_e}{|\bl_e^*|^2} \vert \bu_i - \Pi_{i\rightsquigarrow j} (\bu_j) \vert_{i, j \succ e}^2  =
\sum_{e \in \mathfrak E} \frac{1}{w_e}\vert \coeff{u}_j - e^{- \Img \Omega_{e^*}} \coeff{u}_i \vert_{i, j \succ e}^2.
\end{align}
As mentioned in Sec.~\ref{sec: mesh representation}, the edge length ratio $w_e = |\bl^*|/ |\bl| =  (\cot \angle_{e \prec i} + \cot \angle_{e \prec j}) / 2$ is commonly referred to as the cotangent weight, where $\angle_{e \prec i}$ refers to the corner angle opposite to the edge $e$ at face $i$.

The discrete Bochner Laplacian can be obtained by taking the negated variation of the discrete Dirichlet energy, 
\begin{align}\label{eqn:discrete bochner laplacian}
    \blp: \Gamma_{\mathfrak F}(TM) \rightarrow \Gamma_{\mathfrak F}(TM), \quad A_f (\blp \bu)_f = - \left (\frac{\delta \mathcal{E}}{\delta \bu} \right)_f  = - \Base_f\sum_{e \prec f} \frac{1}{w_e} (\coeff{u}_f - e^{- \Img \Omega_{e^*}} \coeff{u}_i )_{i, f \succ e}.
\end{align}
Here, face $i$ is in the one-ring neighbor of face $f$, and $i$ and $f$ share the edge $e$. 

Similarly, using the vertex-vertex transportation $\Pi_{i \rightsquigarrow j}, i, j \in \mathfrak V$ across edge $e \succ i, j$ (\cf Eqn.~\eqref{eqn: vertex-vertex levi-civita}), we can construct the covariant derivative for a vertex-based nematic field  as follows:
\begin{align}\label{eqn:discrete nematic covariant derivative}
  \cvr^L_{\bl}: \Gamma_{\mathfrak V}(TM) \rightarrow \Gamma_{\mathfrak E}(TM), \quad \vert \bl_e \vert (\cvr_{\bl} [\bq])_e = ([\bq]_i - \Pi_{i\rightsquigarrow j} ([\bq]_j) )_{i, j \prec e} =  \left ( (\coeff{q}_j - e^{- \Img 2 \Omega_{e}} \coeff{q}_i) [\Base]_j \right)_{i, j \prec e}.
\end{align}
The nematic Bochner Laplacian acting on the vertex-based nematic field $[\bq] = \coeff{[q]} [\Base] \in \Gamma_{\mathfrak V}(L)$ is therefore:
\begin{align}\label{eqn:discrete nematic bochner laplacian}
        \blp^L: \Gamma_{\mathfrak V}(L) \rightarrow \Gamma_{\mathfrak V}(L), \quad A_v (\blp [\bq])_v   = - [\Base]_v \sum_{e \prec v} w_e (\coeff{[q]}_v - e^{- \Img 2 \Omega_{e}} \coeff{[q]}_i )_{i, f \prec e},
\end{align}
where vertex $i$ is in the one-ring neighbor of vertex $v$, connected by edge $e$. 

Strictly speaking, there is an ambiguity in the definition of the directional covariant derivative (Eqns.~\eqref{eqn:discrete covariant derivative} and \eqref{eqn:discrete nematic covariant derivative}) because it is equally valid to transport a vector (nematic director) from face (vertex) $j$ to face (vertex) $i$ before applying finite difference. 
However, the discretization of the Dirichlet energy, and thus the Bochner Laplacian (Eqns.~\eqref{eqn:discrete bochner laplacian} and \eqref{eqn:discrete nematic bochner laplacian}), relies only on the norm of the covariant derivative, which remains invariant, $ \vert \Pi_{j \rightsquigarrow i} (\bu_j) -  \bu_i \vert = \vert \bu_j - \Pi_{i \rightsquigarrow j} (\bu_i) \vert$, $ \vert \Pi_{j \rightsquigarrow i} ([\bq]_j) -  [\bq]_i \vert = \vert [\bq]_j - \Pi_{i \rightsquigarrow j} ([\bq]_i) \vert$.

In summary, the intrinsic definition of the discrete Bochner Laplacian emerges naturally from continuous theory of complex line bundles. 
Contrasting the previous covariant divergence (Eqn.~\eqref{eqn: discrete covariant divergence}) that relies on an extrinsic parallel transport achieved through dihedral rotations between the tangent spaces of vertices and faces, the Bochner Laplacian is constructed purely intrinsically by vertex-vertex or face-face parallel transport.
Compared to common approaches that rely on the embedding structure in $\Real^3$ and a projection operator from $\Real^3$ to the manifold, the discrete complex line bundle approach has notably improved efficiency. 
This approach reduces the degree of freedom per element from 3 to 2, and the complex nature furthermore takes advantage of the optimized complex arithmetic handling in established numerical linear algebra libraries.
The corresponding complex-valued Laplace matrix is Hermitian and negative-semidefinite.

\subsection{Curvature, defects, and local-global geometry \label{sec: discrete curvature}}
Here we introduce a discretization of the face Gaussian curvature and an algorithm to detect topological defects and compute their charges. 
This serves as the discrete counterpart to the continuous theory in Sec.~\ref{sec: continuous topology}. 
The construction of curvature and charge leads to discrete analogs of local-to-global theorems including the Poincaré-Hopf and Gauss-Bonnet theorems.

Given a vertex-based $k$-atic field $[\bq]_k: \mathfrak V \rightarrow V_k$, we can utilize vertex-vertex parallel transport $\Pi_{i \rightarrow j} $ (\cf Eqn.~\eqref{eqn: vertex-vertex levi-civita}) with $i, j \in \mathfrak V$ along the incident edge $e \succ i, j$ to measure the finite rotation $ |\bl_e | \eta_e$ (\cf Eqn.~\eqref{eqn: rotation form}) of $[\bq]_k$ along $e$ as follows: 
\begin{align}
    |\bl_e | \eta_e = \arg \left ( \frac{[\bq]_{k, j}}{\Pi_{i \rightsquigarrow j} [\bq]_{k, i}}\right) = \arg \left ( \frac{\coeff{[\bq]}_{k, j}}{ e^{- \Img k \Omega_e } \coeff{[\bq]}_{k, i}}\right).
\end{align}
Assuming the $k$-atic field $[\bq]_k$ is continuous between vertices $i$ and $j$, we choose the principal branch of the $\arg$ function so that $|\bl_e | \eta_e$ lies within $(- \pi, \pi]$.

Recall that in the continuous setting, Gaussian curvature quantifies holonomy, $\int_U K dA = - \int_{\partial U} \omega $ (\cf Sec.~\ref{sec: continuous topology}).
Analogously, discrete Gaussian curvature $K \in C_{\mathfrak F}(M)$ can be represented by closed-loop summation of the primal edge connection $\Omega_e$ as
\begin{align}\label{eqn: face gaussian curvature}
    A_f K_f \coloneqq -\sum_{e \prec f} \Omega_e  \mod 2 \pi \in (- \pi, \pi ].
\end{align}
Selecting branch $(- \pi, \pi]$ for $A_f K_f$ is justified by the small integration domain of face $f$. 
The discrete curvature is also the angle defect at face $f$, given by $A_f K_f = \sum_{v \prec f} \tilde{\angle}_{v}  - \pi$, where $\tilde{\angle}_{v}$ is the geodesic corner angle (\cf Sec.~\ref{sec: mesh setup}). 
The discrete Gauss-Bonnet theorem states, $\sum_{f} A_f K_f = 2 \pi \chi (M)$, where $\chi(M) = |\mathfrak V | - |\mathfrak E| + |\mathfrak F| = 2 - 2 g $ represents the Euler characteristic of the polygonal mesh $M$.

Analogous to Eqn.~\eqref{eqn: charge}, given a $k$-atic field $[\bq]_k$, its topological charge is a fraction-valued function $\mathcal{Z}([\bq]_k): \mathfrak F \rightarrow \mathbb Z /k$ that can be computed for each face $f$ as follows: 
\begin{align}
    \mathcal{Z}_f ([\bq]_k) &= \frac{1}{ 2 \pi k} \sum_{e \prec f}  |\bl_e | \eta_e  + \frac{1}{ 2\pi} A_f K_f.
\end{align}
The charge $\mathcal{Z}_f$ at face $f$ is zero when the adjacent $k-$atic directors $[\bq]_{k, v \prec f}$ are continuous, and nonzero in the presence of defects. 
The charge function satisfies the discrete Poincaré-Hopf theorem exactly, $\sum_f \mathcal{Z}_f ([\bq]_k) = \chi(M)$.

\subsection{Augmented Lagrangian method for Stokes equations \label{sec: augmented lagrangian}}
We utilize a classic AL iteration to solve the incompressible Stokes flow \cite{quarteroni_numerical_1994}. 
Recall that the Stokes equations can be viewed as the stationary condition for a divergence-constrained variational problem (\cf Sec.~\ref{sec: forced stokes flow}). 
The AL approach augmented the original Rayleighian dissipation functional with a quadratic  penalty term on divergence. 
The augmented Rayleighian is given by:
\begin{align}
\begin{split}
       \min_{\bu} \max_p ~ \mathcal{R}_{\mathrm{AL}}  &= 
              \mathcal{R}(\bu, p) +  \frac{k}{2} \llangle \text{div }\bu, \text{div } \bu \rrangle,
\end{split}
\end{align}
where the stationary condition is $(\blp + K + k ~\mathrm{grad} \circ \mathrm{div}) \bu - \text{grad } p + \bff = 0$. 
Here, the penalty coefficient $k$ enforces the incompressibility constraint during the iterative solution process. 
Adjusting the penalty parameter allows effective control over the enforcement of the constraint.
It is worth noting that physically this penalty term is also included in the viscosity term, which provides resistance to expansion and compression using the same mechanism through bulk viscosity. 
Shown in Alg.~\ref{alg:AL stokes}, the solve process involves iteratively solving the modified Stokes equations. 
At each iteration, the velocity and pressure fields are updated until a convergence criterion is met. 
Practically, increasing the value of $k$ makes the fluid very stiff in response to compressibility, which leads to convergence in a small number of iterations. 
However, a high value of $k$ also increases the spectral condition number of the matrix $A = -(\blp + K + k~ \text{grad} \circ \text{div})$, which increases the time per iteration  \cite{quarteroni_numerical_1994}.
An alternative method is to use the Uzawa method combined with pressure projection \cite{quarteroni_numerical_1994}. 
We implement the Poisson solve using the Python linear algebra library \texttt{scipy.sparse}, which utilizes a sparse matrix representation and solves the system using the conjugate gradient method. 
Compared to a direct solver, the conjugate gradient is suitable for medium to large system sizes, as it effectively handles the sparse and symmetric positive-semidefinite nature of the matrices involved.

\begin{algorithm}
\caption{AL subroutine for incompressible Stokes equations}
\label{alg:AL stokes}
\begin{algorithmic}[1]
\Inputs{Penalty coefficient $k$, tolerance $\varepsilon$}
\Initialize{Initial guess $p$}
\While{$\norm{\mathrm{div}~ \bu} > \varepsilon$}
\State {$\bu \gets  \Call{Solve}{
    -(\blp + K + k ~\mathrm{grad} \circ \mathrm{div}) \bu =  \bff - \text{grad } p } $} \Comment{Poisson solve ($\texttt{A x = b}$)}
\State {$ p \gets p + k ~\mathrm{div}~ \bu $}
\EndWhile
\\
\If{exist $N$ distinct $\Bomega_i$}\Comment{Remove the Killing component}
\For{$i = 1$ to $N$}
\State {$ \bu \gets \bu - \llangle \bu, \Bomega_i \rrangle \Bomega_i $}
\EndFor
\EndIf 
\end{algorithmic}
\end{algorithm}

As discussed in Sec.~\ref{sec: forced stokes flow}, when dealing with the Stokes equations in symmetric geometries that possess a Killing vector field, the solution is not unique. 
Despite this, iterative methods such as the conjugate gradient method can still converge to one feasible solution. 
To ensure that we obtain continuous solution during time evolution, we consistently select the solution with the minimal $L_2$ norm by projecting out components that lie in the space of the Killing vector fields. 
To achieve this, we precompute the set of Killing bases of the geometry, denoted as $\Bomega_i$, before we begin the evolution of the active nematic system. 
The bases are determined by solving the nonvanishing vector field that satisfies $(\blp + K + \grad \circ \div) \Bomega_i = 0$.
In practice, we solve this equation computationally through the smallest eigenvalue problem and extract eigenfunctions with near-zero eigenvalues. 
Here each Killing basis is normalized such that $\llangle \Bomega_i , \Bomega_j \rrangle = \delta_{ij}$. 
As outlined in Alg.~\ref{alg:AL stokes}, we project out components spanned by these bases after the AL iteration. 
This protocol is carried out in the Sec.~\ref{sec:results} for examples on analytical shapes.
Note that the number of distinct Killing vector fields of a given geometry is often apparent from its symmetry prior to performing the actual computation.
For example, a sphere exhibits three distinct Killing vector fields, and a torus displays one,  corresponding respectively to their modes of rotational symmetries.

\subsection{Implementation \label{sec:implementation}}
Along with the theoretical and algorithmic development, we also provide an implementation using SideFx Houdini, a freely-accessible software under its educational license.  
Our implementation is accessible through a public repository on Github: \url{https://github.com/CunchengZhu/Riemannian-active-nematics-2024.git}. 
SideFx Houdini is primarily known as a 3D animation and special effects software. 
However, it can be effectively utilized for custom algorithm implementation by leveraging its half-edge data structure of triangular meshes, as well as its \texttt{VEX} and Python programming capabilities. 
\texttt{VEX} is a C-like shader language that operates in a vectorized, multithreaded manner. 
The Python module enables access to the numerical linear algebra libraries \texttt{Scipy}. 
Additionally, the software provides visualization and basic geometric processing features such as mesh parametrization and interpolation to facilitate rapid prototyping.

\section{Results and discussion \label{sec:results}}
In the preceding sections, we established the continuous formulation, its corresponding discretization, and the numerical methods employed. 
In this section, we demonstrate the robustness of the method and its potential in addressing various biophysically relevant scenarios.
First, we conduct a convergence analysis spanning the various subroutines of our methodology. 
This analysis not only attests to the method's accuracy in approximating the smooth solution but also illustrates the performance of the algorithms. 
In addition, we explore the efficiency and scalability of the methods by presenting the computational-time scaling characteristics of the major components within the framework. 
We have chosen to benchmark these methods on simple geometries, where a mix of numerical experiments and analytical comparisons are possible. 
However, the application extends beyond idealized geometries. 
To demonstrate the generality of these methods on arbitrary geometries and topologies, we apply them to solve the full nematodynamics equations in biophysically relevant examples. 
In this section, through the error analysis, scalability metrics, and examples, we expect to demonstrate the reliability and versatility of the method.

\subsection{Convergence analysis\label{sec: convergence}}
We perform convergence tests for our method using a combination of analytical results and numerical experiments. 
We break down the method into different steps, including the Lie advection of the nematic field and the solution of the surface Stokes equations.

\begin{figure}[t]
    \centering
    \includegraphics[width= 5.5 in]{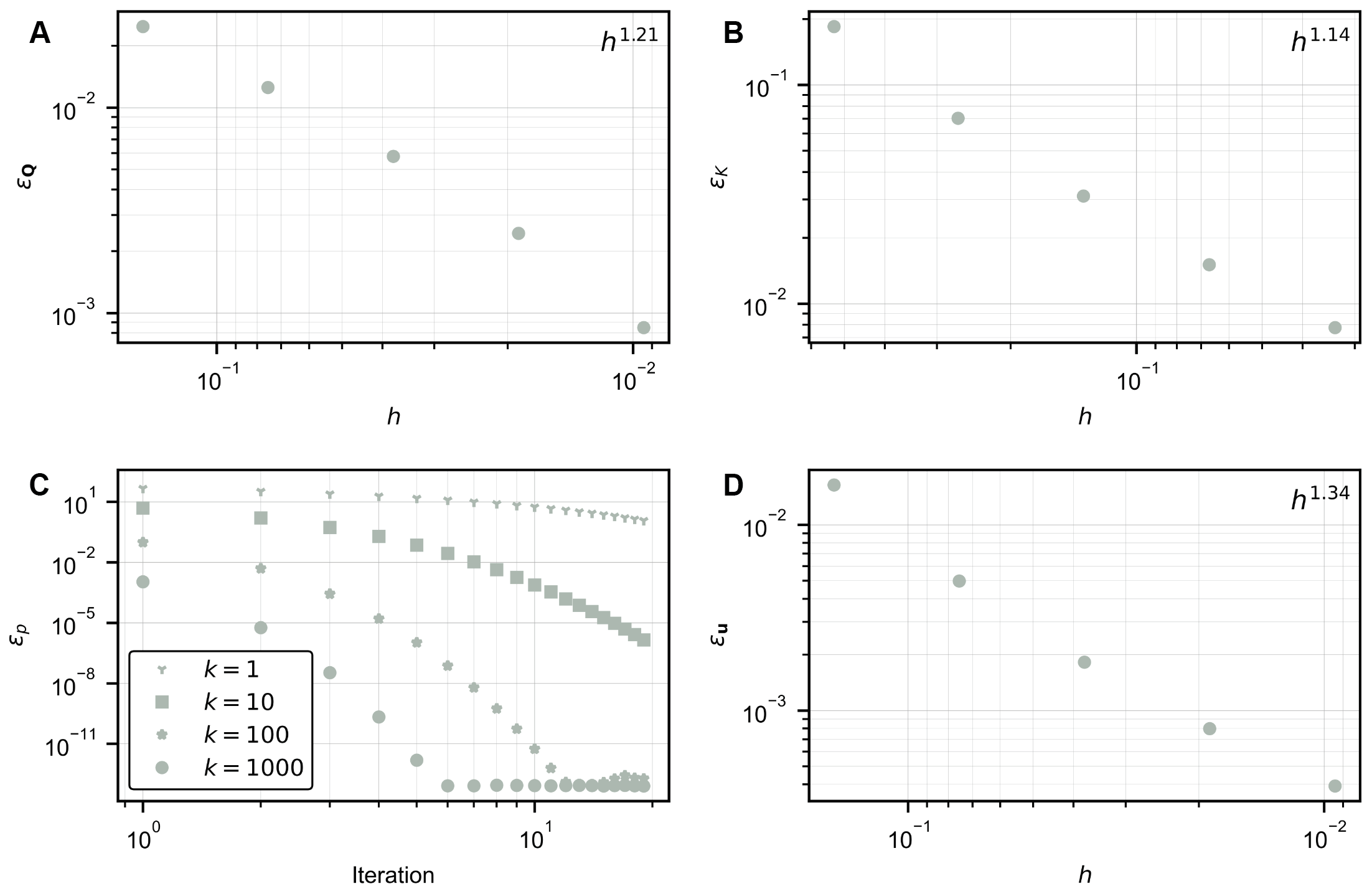}
    \caption{Convergence test of different aspects of the method: A) Nematic sL advection error with mesh refinement; B) Gaussian curvature error with mesh refinement; C) Area dilation of the fluid velocity solution after each AL pressure iteration; and D) Error of Stokes equations solution with mesh refinement.  }
    \label{fig: convergence}
\end{figure} 

We start by testing the spatial convergence of the nematic semi-Lagrangian advection on a unit sphere. 
To do this, we parameterize the unit sphere using spherical coordinates, with polar angle $\theta\in[0,\pi]$ and azimuthal angle $\varphi\in[0,2\pi)$, which results in the spherical vector bases, $\be_{\theta} \coloneqq \partial_\theta / |\partial_\theta| $ and $\be_\varphi \coloneqq \partial_\varphi / |\partial_\varphi |$.
We prescribe an analytical nematic field along the unit polar basis vector $\be_\theta$, which creates two $+1$ defects at the poles of the sphere. 
We also prescribe an advecting velocity $\bu$, which is defined using solenoidal vector spherical harmonics with $l = 2$, $m = 0$ (\ie $\bu = \Bphi_{20} =  (3/2) \sqrt{5/\pi} \sin \theta \cos \theta ~\be_\theta$).
We then advect the nematic field for one time step using step size $\tau = 0.1$ to measure the local error. 
We use the numerical result obtained from the finest mesh tested as the ground truth, denoted as $\overline{\bQ}$, and compare results from coarser meshes using the $L_2$ relative error norm, $\varepsilon_{\bQ} \coloneqq \norm{\bQ - \overline{\bQ}} / \norm{\overline{\bQ}} $. 
As shown in Fig.~\ref{fig: convergence}A, the error decreases approximately linearly with mesh refinement, following $\varepsilon_{\bQ} \propto h^{1.21}$, where $h$ represents the average edge length of the mesh.
Note that here we omitted the unit-length normalization step in Alg.~\ref{alg:outline} and directly compared the embedded $\Real^{3 \times 3}$ matrix representation of $\bQ$.

The Stokes momentum equation involves the estimation of Gaussian curvature $K$.
To test this, we use a torus for which the analytical expression of Gaussian curvature is known, with a value that varies from negative to positive. 
We consider a torus with a radius of revolution $R$, and a radius of cross-section $r$, for which the Gaussian curvature can be parametrized using the polar coordinate as $\overline{K}(\theta) = {\cos \theta}/ {(r R + r^2 \cos \theta)}$. 
We calculate the face-based Gaussian curvature $K$ using Eqn.~\eqref{eqn: face gaussian curvature} and compare it to the analytical results using a relative $L_2$ error, denoted as $\varepsilon_K \coloneqq \| K - \overline{K} \| / \| \overline{K}  \|$, where $\overline{K}(\theta)$ is computed at the center of the face. 
Fig.~\ref{fig: convergence}B demonstrates that the error decreases linearly with mesh refinement, following $\Bvarepsilon_{\bQ} \propto h^{1.14}$.

We also evaluate the procedure of the AL that updates the pressure $p$ to enforce the incompressibility constraint. 
This involves measuring the pressure error norm that is defined by the the norm of area dilation, $\varepsilon_p \coloneqq | \mathrm{div} ~\bu | $. 
Fig.~\ref{fig: convergence}C illustrates that increasing the penalty coefficients $k$ results in faster convergence to a divergence-free solution. 
However, this comes at the cost of making the matrix inversion ill-conditioned, which increases the time per iteration.

Finally, we compare the numerical solution of the Stokes equations with its analytical counterpart.
We construct the forcing term $\bff$ in the momentum equation using the divergence-free spherical harmonics basis function, which is an eigenfunction of the Stokes equations. 
This approach allows us to derive an analytical solution $\bar \bu$ for the surface flow on a sphere. 
We chose $\bff$ to be a mix of two eigenmodes, $\bff = \Bphi_{21} + \Bphi_{20}$, and generated numerical solutions $\bu$ across different mesh resolutions. 
The error, as illustrated in Fig.~\ref{fig: convergence}D, decreases linearly with the refinement of the mesh, following the relation $\varepsilon_{\bu} \coloneqq \| \bu - \overline{\bu} \| / \| \overline{\bu} \| \propto h^{1.34}$.

\subsection{Computational time complexity\label{sec: time complexity}}
\begin{figure}[t]
    \centering
    \includegraphics[width= 3 in]{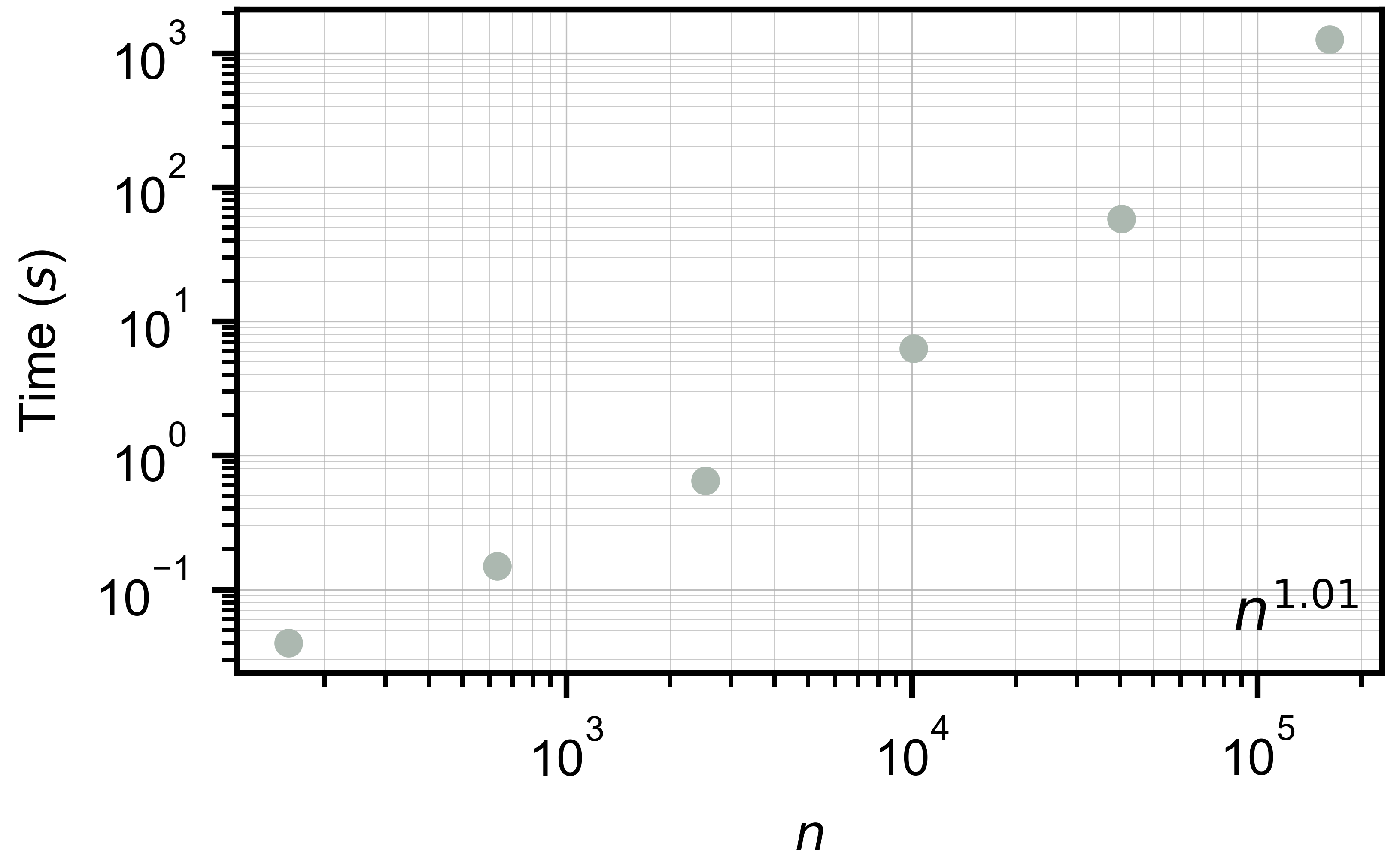}
    \caption{Computational time complexity of AL iterations for solving the Stokes equations: the linear solve scales linearly with the system size $n$, or the total number of faces in the mesh. }
    \label{fig:scaling}
\end{figure}

The Lie advection of the nematic field is explicit. 
The computation is also local and parallelizable, which results in minimal computational cost.
The principal cost of the algorithm stems from the assembly of differential matrices and the subsequent linear solve associated with the AL iteration. 
The assembly of matrices, however, needs to be precomputed only once. 
Thus, the predominant computation load is attributed to the Poisson solve $\texttt{A x = b}$ for the Stokes equations in each AL iteration.
In this study, $\texttt{A}$ is a symmetric positive definite Laplacian matrix. 
Efficient solutions can be attained by standard sparse linear algebra libraries. 
For smaller system sizes, we employ the direct solve in Python's $\texttt{Scipy.sparse}$ library. 
As the system size increases, an iterative method such as conjugate gradient method becomes more pertinent.  
Empirically, as observed in Fig.~\ref{fig:scaling}, the computation time for the AL iteration exhibits linear scaling with the system size $n$, where $n$ is the total number of faces in the mesh. 

These tests were conducted on a Dell 8940 Desktop equipped with an Intel CPU i9-11900K operating at 3.50 GHz, 64 GB of RAM. 
In instances where a finer mesh is employed, comprising approximately 10,000 vertices—akin to the examples illustrated in the paper—the computation at each time step requires approximately 2 minutes. 

\subsection{Nematodynamics on arbitrary surfaces\label{sec: numerical experiments}}
To validate our proposed methodology and demonstrate its versatility across a spectrum of shapes with varying geometry and topology, we conducted a series of numerical experiments. 
These experiments encompassed analytical shapes such as spheres and genus-1 tori, as well as organic shapes characterized by arbitrary curvature distributions.
As previously mentioned in Secs.~\ref{sec: forced stokes flow} and \ref{sec: augmented lagrangian}, the Stokes equations have Killing vector fields as their kernel.
On spheres and tori where such a Killing vector field exists, we carried out a projection to remove the Killing component and focus on a specialized solution with a minimal $L_2$ norm, $\|\bu \|$.
We present each example under two distinct levels of activity, distinguished by disparate Péclet numbers ($\Pe = 1$ and $\Pe = 10^4$) and  effectively showcasing both regular solutions and solutions in the regime of active turbulence.

\begin{figure}[t]
    \centering
    \includegraphics[width= \textwidth]{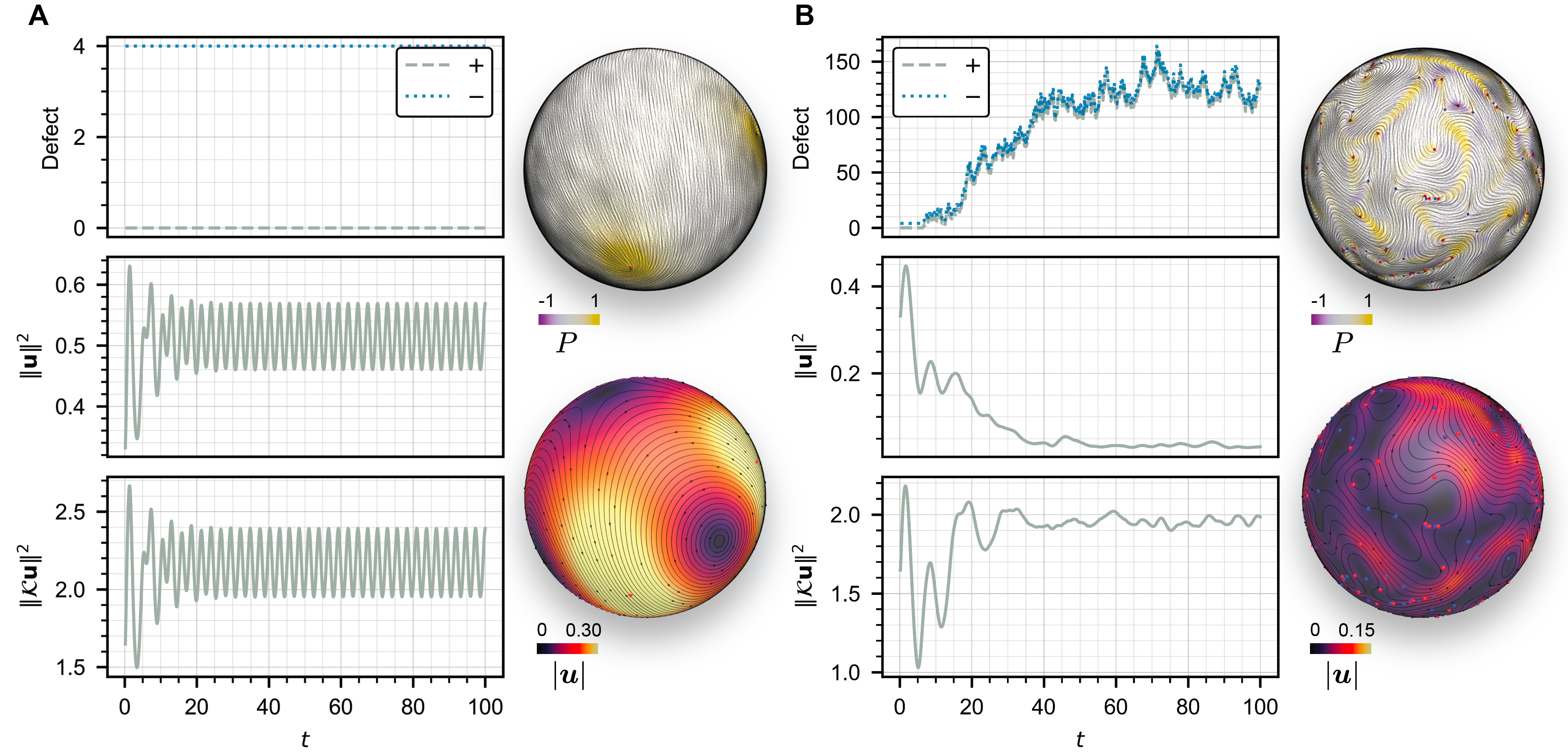}\vspace{-0.4cm}
    \caption{
     Nematodynamics on a sphere under two levels of activity, as captured by the P\'eclet number: A) low activity ($\Pe = 1$), and B) high activity ($\Pe = 10^4$). 
     The animated time series simulation is available on Github: \url{https://github.com/CunchengZhu/Riemannian-active-nematics-2024/blob/main/sphere_compressed.mp4}.
     From top to bottom, the time plots show: the total numbers of $+1/2$ and $-1/2$ topological defects of the nematic field, the $L_2$ norm squared $\|\bu\|^2$ of the fluid velocity, and the enstropy $\| \mathcal{K} \bu \|^2$ of the surface flow. 
     Snapshots of the system are captured at the conclusion of each set of simulations. 
     The upper snapshots display the nematic field, color-mapped by the local power density exerted by the nematic onto the fluid, expressed as $P = -\langle \bu, \mathrm{div}^{\cvr} \bQ \rangle $.
     The lower snapshots visualize the velocity field, color-mapped by its magnitude~$|\bu|$.}
    \label{fig: sphere}
\end{figure}

To characterize the nematodynamics, we focused on recording key metrics, including the count of positive and negative defects (of charge $\pm 1/2$), the $L_2$ norm squared of the flow velocity, $\| \bu \|^2$, and the entrosphy of the fluid, $\|\mathcal{K} \bu \|^2$.
The entrosphy quantifies the magnitude of viscous dissipation and serves to assess the sharp features and chaotic behavior of the flow field.
For each case at each Péclet number, we appended a visual snapshot of the nematodynamics simulation at the end of each set of simulations. 
Taking Fig.~\ref{fig: sphere} as an example, the top snapshot displays the texture of the nematic field, with positive defects shown in red and negative defects in blue.
The bottom snapshot illustrates the streamlines of the fluid flow field, color-mapped to represent velocity magnitude, $|\bu|$. 
The colormap on the nematic field represents the local power density exerted by the active nematic on the fluid, expressed as $ P = -\langle \bu, \mathrm{div}^{\cvr} \bQ \rangle$.
Each positive half defect is accompanied by a self-propelling fluid jet and a vortex pair, and is surrounded by a region of positive power $P$. 
In contrast, negative half defects exhibit 3-atic symmetry and generate an index-2 flow field that is self-counteracting and remains static \cite{giomi2014defect}, contributing little energy to the fluid on large scales.

In Fig.~\ref{fig: sphere}A, we examine dynamics on a sphere at Péclet number $\Pe = 1$, corresponding to a low activity level. 
Here, we observe a fixed number of four $+1/2$ defects exhibiting periodic orbiting motions--—a phenomenon supported by previous empirical studies and simulations \cite{keber_topology_2014, firouznia2024self}.
In Fig.~\ref{fig: sphere}B, we increase the activity level to $\Pe = 10^{4}$, leading to the emergence of active turbulence.
Active turbulence is characterized by unsteady and chaotic dynamics driven by the spontaneous formation and annihilation of topological defect pairs.
Demonstrated by Fig.~\ref{fig: sphere}B and consistent with the existing literature, the onset of turbulence is characterized by the appearance of large bending deformation walls that separate aligned nematic regions, the creation and annihilation of nematic defect pairs, and the separation of positive defects from negative defects due to self-propulsion of the positive \cite{thampi_active_2016}. 
As the total number of topological defects increases, intricate flow patterns develop at smaller length scales. 
This observation is supported by the fluctuating but steady enstrophy and a rapid decrease in the overall velocity magnitude in the time series plots.

\begin{figure}[t]
    \centering
    \includegraphics[width= \textwidth]{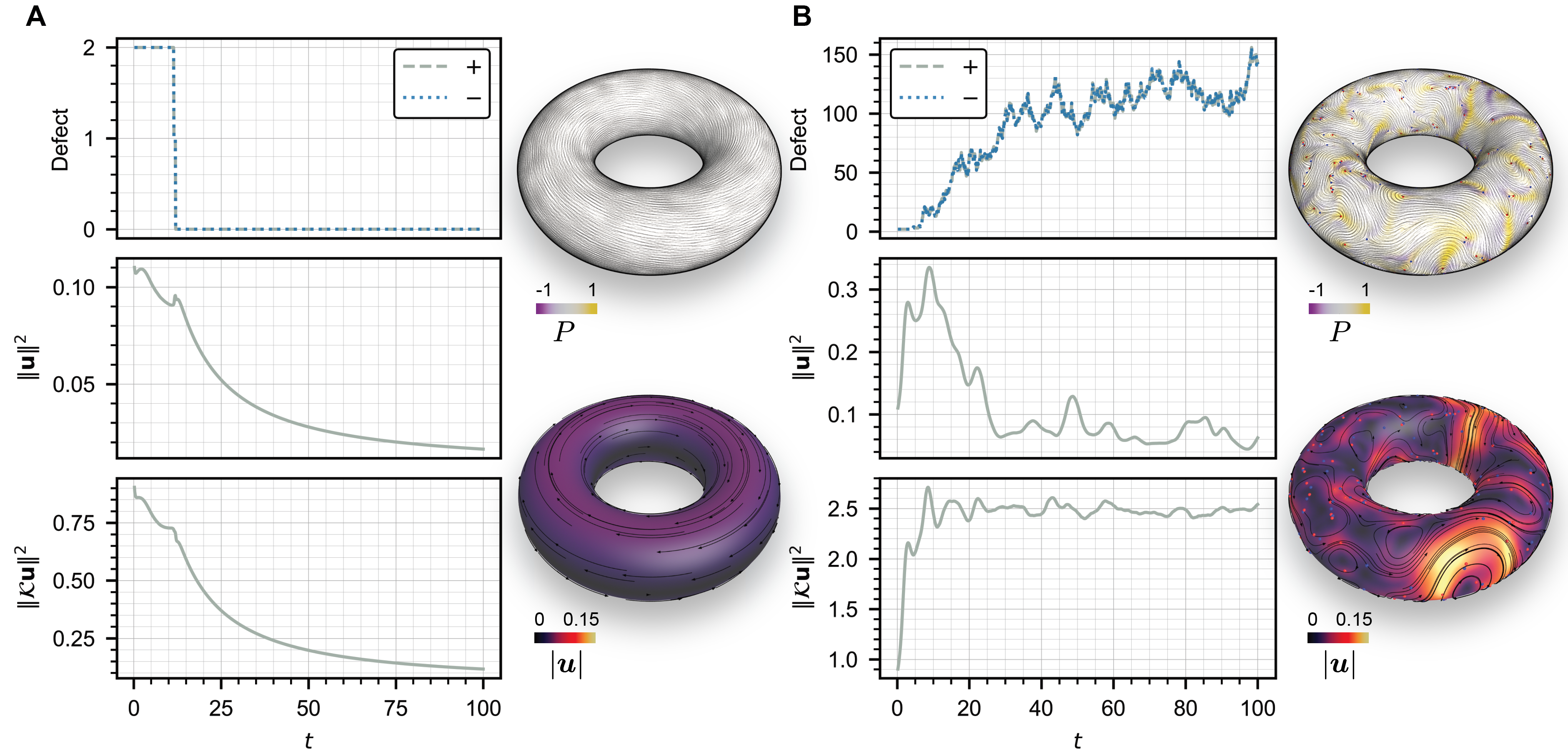}\vspace{-0.4cm}
        \caption{Nematodynamics on a torus at $\Pe = 1$ and $\Pe = 10^4$. 
        The animated time series simulations is available on Github: \url{https://github.com/CunchengZhu/Riemannian-active-nematics-2024/blob/main/torus_compressed.mp4}. See Fig. \ref{fig: sphere} for a detailed caption.}
    \label{fig: torus}
\end{figure}

As our method is able to accommodate different topologies, we expand our exploration of nematodynamics to a genus one torus, characterized by a zero Euler characteristic. 
The vanishing Euler characteristic allows for a defect-free configuration at lower activity levels, as the nematic diffusion gradually merges two pairs of positive and negative defects (\cf Fig.~\ref{fig: torus}A). 
A key observation is that the equilibrium defect-free nematic field on the torus displays rotational symmetry around its vertical axis, while maintaining a constant angle with the lines of toroidal coordinates. 
Consequently, the resulting Stokes flow also exhibits this rotational symmetry, with the toroidal midline acting as a shear layer. 
This flow pattern, in conjunction with diffusion, stabilizes the nematic field, allowing the system to reach a steady state.
In Fig.~\ref{fig: torus}B, when the Péclet number increases, the system transitions into a state of active turbulence through the injection of energy by topological defects, similar to what occurs on a sphere. 
Once again, we note that enstropy reaches a plateau soon after the initial development of defect bindings, while the velocity norm decreases as vortices form on increasingly smaller length scales. 
\begin{figure}[t]
    \centering
    \includegraphics[width= \textwidth]{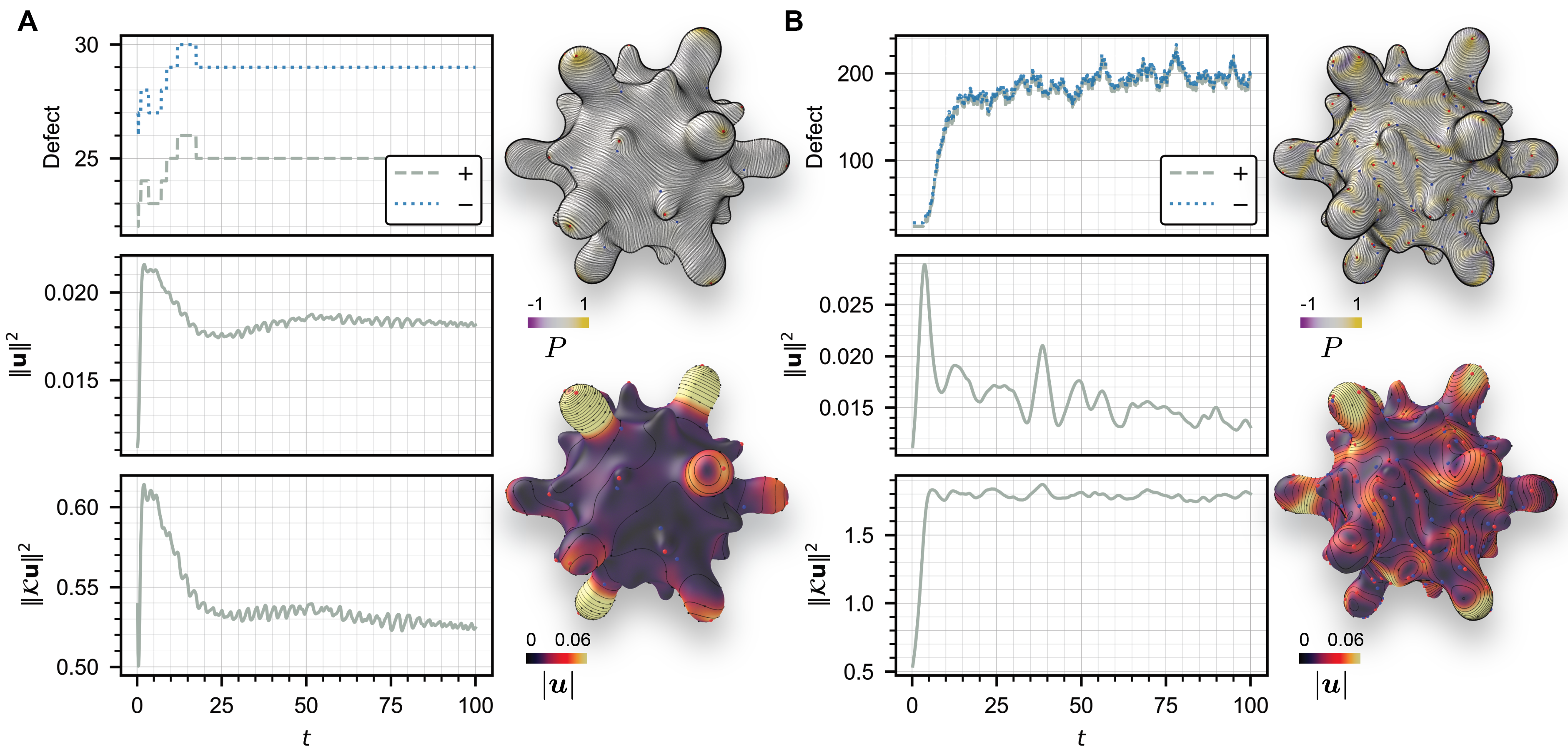}\vspace{-0.4cm}
    \caption{Nematodynamics on a topological sphere of varying curvature at $\Pe = 1$ and $\Pe = 10^4$. 
    The animated time series simulations is available on Github: \url{https://github.com/CunchengZhu/Riemannian-active-nematics-2024/blob/main/spike_compressed.mp4}.
    See Fig. \ref{fig: sphere} for a detailed caption.}
    \label{fig: virus}
\end{figure}

In addition to standard analytical shapes, we applied our method to a topologically spherical membrane structure with an arbitrary curvature distribution (\cf Fig.~\ref{fig: virus}A). 
This geometry was generated through a simulation based on the Helfrich bending energy under the influence of an osmotic shock \cite{zhu2022mem3dg}.
Even under low-activity conditions where nematic ordering relaxation prevails, a relatively large total number of defects persists. 
Notably, these defects tend to cluster in areas with like-sign curvature; that is, positive defects gather at the protrusions with positive curvature, while negative defects accumulate in negatively curved valleys.
When two positive defects are present at a protrusion, they create strong circulating flows at the tips. 
These flows cause the defects to orbit periodically around the protrusions. 
This interaction at the protrusion is isolated from the rest of the domain, which remains almost motionless, thereby creating a limit cycle in the system. 
The distinct behaviors of active nematics in areas of positive and negative curvature suggests a possible detailed comparative study on active nematic systems in hyperbolic and elliptic geometries as a followup work. 
More broadly, the division of the flow field into areas of positive and negative curvature has implications for understanding how mixing and coherent structures are influenced by the shape of the domain.
In the high activity regime, there is a notable difference in flow velocities when comparing conditions on a complex-shaped object to the sphere case. 
On a sphere, the interaction of topological defects creates complex flow patterns on small length scales, which tends to reduce the characteristic magnitude of the fluid velocity. 
However, as seen in Fig.~\ref{fig: virus}B, for the complex shape, the magnitude of velocities remains relatively consistent whether in a steady state at low $\Pe$ or in active turbulence at high $\Pe$. 
As $\Pe$ increases, active turbulence disrupts the stable, coherent structures that are present at low $\Pe$, leading to a state with more efficient mixing by the flow. 
Specifically, this results in a decrease in velocity at the protrusions and an increase in velocity in the valleys of the manifold.

\section{Conclusion \label{sec:conclusion}}
In this study, we formulate a minimal model for the dynamics of an active nematic fluid on a Riemannian 2-manifold. 
This model is cast based on a coordinate-free differential-geometric language and on the theory of the complex line bundle, which can naturally be generalized to curved spaces. 
We use the complex line to establish the 2D $k$-atic equivalence and model nematic advection through generalized Lie derivatives, laying the groundwork for extending the model to $k$-atic hydrodynamics. 
The Levi-Civita connection and its curvature are introduced within the framework of the complex line bundle. 
We show the coupling of the fluid's viscous stress with the curvature through the covariant Dolbeault $\delbar $ operator.
Numerically, our method evolves the active nematic system on a triangular mesh with arbitrary shape and topology. 
This is done by adhering to the discrete analogue of the continuous theory.
We construct the Hermitian Bochner Laplacian matrix based on the variational principle and develop a generalized semi-Lagrangian scheme according to the Lie advection. 
This method maintains robustness and efficiency across various flow regimes, from low to high activity levels.

In future work, we anticipate detailed studies that examine the relationship between the complex number representation and the well-established $\bQ$-tensor theory, including their generalizations to $k$-atic fields. 
We also acknowledge that our minimal Riemannian nematodynamics model does not account for the possible influence of extrinsic geometry on nematic and fluid dynamics, which may be significant in various applications. 
In subsequent work that will be detailed elsewhere, we explore the surface Stokes flow of evolving manifolds, where fluid flow induces extrinsic deformations of the surface.
Depending on the specific application, it may also be interesting to include factors such as membrane bending rigidity, interactions with bulk fluid, and reorientation of nematic molecules along the extrinsic principal curvatures.

In summary, our integration of geometric language and complex manifold theory into the study of active nematics sets the stage for further theoretical and computational developments in this area. 
With further post-processing and analysis, we expect that the computational tool developed here will also be instrumental for specialists seeking deeper understanding of active nematics influenced by geometric aspects in specific systems.
We hope that the incorporation of realistic geometric and topological features in the model can further narrow the gap between theoretical predictions and experimental findings.

\section*{Acknowledgements\label{sec: acknowledgements}}
C.Z. thanks M. Firouznia and Y. Chen for their helpful references and many discussions on various topics covered in this study. 
D.S. acknowledges funding from National Science Foundation Grant No. DMS-2153520. 
A.C. acknowledges funding from National Science Foundation CAREER Award 2239062.
Additional support was provided by SideFX software.
\appendix
\renewcommand{\theequation}{\thesection.\arabic{equation}}

\section{Derivation of $\bar \partial^{ \nabla *} \bar \partial^{ \nabla}$ using complex differential form}\label{sec: complex derivation}
\setcounter{equation}{0}

In this section, we derive the viscosity Laplacian, $-2 \bar \partial^{\cvr *} \delbar  = \blp + K: \Gamma(TM) \rightarrow \Gamma(TM)$, through the use of complex differential forms, assuming a basic knowledge of exterior calculus.
Consider a Hermitian complex line bundle \(L\) over \(M\) with connection \(\nabla\colon\Gamma(L)\to\Omega^1(M;L)\), where $\Omega^k(M;L)$ represents the space of $L$-valued $k$-forms. %
The curvature 2-form for the connection, $R^\nabla\in\Omega^2(M;\Complex)$, is defined by
\begin{align}
R^\nabla \llbracket \bv,\bw\rrbracket \Bpsi \coloneqq (d^\nabla\nabla \Bpsi)\llbracket \bv,\bw\rrbracket 
= \nabla_{\bv}\nabla_{\bw} \Bpsi - \nabla_{\bw}\nabla_{\bv} \Bpsi - \cvr_{[\bv,\bw]} \Bpsi.
\end{align}
Note that when $L = TM$ and $\nabla$ is the Levi-Civita connection, $R^\nabla$ correlates directly with the Gaussian curvature $K$ as $ R^\nabla = - \Img K\star 1$.

Note that for a 1-form $\alpha \in \Omega^1(M; L)$ on a 2-manifold, the relation $\alpha \pair {\Img \bv} = -\star\alpha \pair{v}$ holds, where $\star$ denotes the Hodge star operator.  
The anti-holomorphic derivative, \(\bar\partial^\nabla \Bpsi \), \( \Bpsi \in\Gamma(L)\), induced by the connection \(\nabla\), can be expressed as  
\begin{align}
    \bar \partial^\nabla \Bpsi \coloneqq \frac{1}{2} (\nabla \Bpsi + \Img \nabla_{\Img (\cdot)} \Bpsi) = \frac{1}{2} (\nabla \Bpsi - \Img \star \nabla \Bpsi).
\end{align}
The anti-holomoprhic derivative allows us to construct the anti-holomorphic energy as follows:
\begin{align}\label{eqn: complex viscosity}
\begin{split}
      2 \langle \delbar\Bpsi , \delbar\Bpsi  \rangle \star 1 
      &= \frac{1}{2} \langle \cvr \Bpsi - \Img \star \cvr \Bpsi ,   \cvr \Bpsi - \Img \star \cvr \Bpsi \rangle \star 1  \\
      & = \frac{1}{2} \left ( \langle \cvr \Bpsi , \cvr \Bpsi  \rangle  \star 1 + \langle \Img \star \cvr\Bpsi , \Img \star \cvr\Bpsi  \rangle  \star 1 - 2 \langle \Img \star \cvr\Bpsi , \cvr\Bpsi  \rangle \star 1  \right)\\ 
  &= \langle \cvr\Bpsi , \cvr\Bpsi  \rangle  \star 1 +\langle \cvr\Bpsi  {\wedge} \Img \cvr\Bpsi  \rangle \\
   & =\langle \cvr\Bpsi , \cvr\Bpsi  \rangle  \star 1 +   d\langle\Bpsi \wedge \Img \cvr\Bpsi  \rangle - \langle\Bpsi  \wedge \Img d^{\cvr} \cvr \Bpsi \rangle \\
      & = \langle \cvr\Bpsi , \cvr\Bpsi  \rangle  \star 1 - \langle\Bpsi , \star \Img d^{\cvr} \cvr\Bpsi  \rangle = \langle \cvr\Bpsi , \cvr\Bpsi  \rangle  \star 1 - \langle\Bpsi ,  \Img \star R^\nabla \Bpsi  \rangle.
\end{split}
\end{align}
By taking a variation with respect to $\Bpsi$ on the final expression, we obtain the operator $-2 \bar \partial^{\cvr *} \delbar  = \Delta + \Img \star R^\nabla : \Gamma(L) \rightarrow \Gamma(L)$.
When the complex line bundle is specialized to the tangent bundle, $L = TM$, the operator is the viscosity Laplacian, $2 \bar \partial^{\cvr*} \delbar  = \Delta + K: \Gamma(TM) \rightarrow \Gamma(TM) $.

\bibliographystyle{model1-num-names}
\bibliography{ref}
\end{document}